\documentclass[a4paper,11pt,final]{article}

\usepackage{authblk}

\usepackage{graphics}
\usepackage{graphicx}
\usepackage{epsfig}
\usepackage{amssymb}
\usepackage{amsthm}

\usepackage{amsmath}
\date{}

\begin{document}

\title{Stochastic foundations of
undulatory transport phenomena:  Generalized Poisson-Kac processes - Part II
 Irreversibility, Norms and Entropies}

\author[1]{Massimiliano Giona$^*$}
\author[2]{Antonio Brasiello}
\author[3]{Silvestro Crescitelli}
\affil[1]{Dipartimento di Ingegneria Chimica DICMA
Facolt\`{a} di Ingegneria, La Sapienza Universit\`{a} di Roma
via Eudossiana 18, 00184, Roma, Italy  \authorcr
$^*$  Email: massimiliano.giona@uniroma1.it}

\affil[2]{Dipartimento di Ingegneria Industriale
Universit\`{a} degli Studi di Salerno
via Giovanni Paolo II 132, 84084 Fisciano (SA), Italy}

\affil[3]{Dipartimento di Ingegneria Chimica,
 dei Materiali e della Produzione Industriale
Universit\`{a} degli Studi di Napoli ``Federico II''
piazzale Tecchio 80, 80125 Napoli, Italy}
\maketitle

\begin{abstract}
In this second part, we analyze the dissipation properties of Generalized
Poisson-Kac (GPK) processes, considering the decay of suitable $L^2$-norms 
and the definition of entropy functions. In both cases,
consistent energy dissipation and entropy functions depend on
the whole system of primitive statistical variables, the
partial probability density functions $\{ p_\alpha({\bf x},t) \}_{\alpha=1}^N$,
while the corresponding energy dissipation and entropy
functions based on the overall probability density $p({\bf x},t)$
do not satisfy monotonicity requirements as a function of time.
Examples from chaotic advection (standard map coupled to
stochastic GPK processes) illustrate this phenomenon.
Some complementary physical issues are also addressed: the ergodicity
breaking in the presence of attractive potentials, and the
use of GPK perturbations to mollify stochastic field equations.
\end{abstract}

\section{Introduction}
\label{sec_1}

The setting of Generalized Poisson-Kac processes (GPK, for short)
has been addressed in detail in part I \cite{part1}, explaining
their physical motivation, as a generalization of
the Kac's paradigm of stochastic processes possessing finite
propagation velocity \cite{kac}, and their structural properties
with particular emphasis on the Kac limit. In point of fact,
the Kac limit represents a form of asymptotic consistency
of GPK stochastic differential equations with respect to 
classical Langevin equation driven by Wiener processes (which
 can be also referred to as the {\em Brownian motion
consistency of GPK processes}).
We refer to part I for the notation and the basic properties
of GPK dynamics that are not reviewed here to avoid repetition.

In this second part of the work we focus on the characterization
of irreversibility in GPK dynamics, essentially grounded
on the definition of suitable $L^2$-norms (energy dissipation
functions) based on the system of the partial probability
density functions   $p_\alpha({\bf x},t)$, $\alpha=1,\dots,N$,
and possessing a monotonic decay in time, and 
of proper entropy functions for GPK processes. Section \ref{sec_2}
is entirely dedicated to this issue.
In both cases, dissipation and entropy functions 
can be defined for GPK processes of increasing structural
complexity and depend on the whole system of partial probability
waves. Starting from the simplest cases, we consider transitionally
symmetric GPK processes  \cite{part1} and extend the analysis to
the transitionally  non-symmetric case, admitting
relativistic implications.

Moreover, energy dissipation and entropy functions constructed
solely upon the knowledge of the overall probability density 
$p({\bf x},t)= \sum_{\alpha=1}^N p_\alpha({\bf x},t)$
do not satisfy the requirement of monotonicity in time, and
consequently are thermodynamically inconsistent.
This is a first important physical indication on the
fact that the primitive statistical description of GPK processes
is entirely based on the whole set of partial probability waves
and cannot be reduced to the coarser description based
on the overall probability density function $p({\bf x},t)$
and its associated probability density flux ${\bf J}_d({\bf x},t)$,
 see part I.
This result is fully consistent with the underlying hypothesis
of extended thermodynamic theories \cite{ext1,ext2,ext3},
and indicates that GPK processes are the simplest candidate
for  the microdynamic equations of motion in extended theory of
far-from-equilibrium phenomena. This issue is further elaborated in
part III \cite{part3}.

A physically meaningful example illustrating these
properties refers to  chaotic advection of tracer particles
in the presence of a stochastic perturbation (diffusion),
modelled as a GPK process. As a prototypical model flow we
consider the continuous-time flow associated with
the standard map \cite{sm1,sm2} 
on the two-dimensional torus (Section \ref{sec_3}).

Finally Section \ref{sec_4} addresses some auxiliary physical properties
of GPK processes: (i) the use of GPK perturbations to mollify
stochastic field equations (stochastic partial differential
equations) \cite{stocafield1,stocafield2}, and (ii) the
occurrence of ergodicity breaking in the
presence of attractive potentials.
The analysis of one-dimensional models addressed in \cite{giona_epl}
is briefly reviewed, and the theory is extended to higher-dimensional
systems.

\section{Norm dynamics, fluxes and entropy}
\label{sec_2}

The definition and evolution of suitable $L^2$-norms
accounting for the dissipation induced by stochasticity is strictly connected with the setting of 
a proper entropy function.
Both the definition of an energy-dissipation
function, based on the $L^2$-norms of the characteristic
partial probability waves, and that of an entropy function depend
not solely on the overall probability density function $p({\bf x},t)$, as for
Wiener-driven Langevin equations, but also on other dynamic
quantities describing the process. These quantities
are simply the diffusive flux in the one-dimensional
Poisson-Kac process, or a combination of fluxes and other
auxiliary quantities in the general GPK case. The analysis of the
GPK case reveals that the use of primitive statistical
quantities, i.e., of the partial probability waves, provides
the simplest and physically meaningful description
of dissipation.

The exposition is organized as follows. To begin
with, the one-dimensional Poisson-Kac process in the absence of
deterministic biasing fields is addressed.  Subsequently,
the case of GPK processes is thoroughly treated,
starting from simpler cases up to
the more general case of transitionally non-symmetric
GPK dynamics. The latter case is relevant in
connection with the relativistic setting 
of GPK dynamics, and with their transformation properties
under a Lorentz boost.

\subsection{One-dimensional Poisson-Kac diffusion}
\label{sec_2_1}

Consider the one-dimensional Poisson-Kac process 
in the absence of a deterministic bias, i.e.,
 $d x(t) =  b \, (-1)^{\chi(t)}$, where the Poisson process $\chi(t)$ is
characterized by the transition rate $\lambda>0$ .
In the Kac limit, it corresponds  to a purely diffusive Brownian motion,
possessing an effective diffusivity $D_{\rm eff}=b^2/2 \lambda$.
Its statistical description involves the two partial probability
waves $p^{\pm}(x,t)$, satisfying the hyperbolic system
of equation
\begin{equation}
\partial_t p^\pm(x,t)= \mp b \, \partial_x p^{\pm}(x,t) \mp \lambda
\left [ p^+(x,t) -p^-(x,t) \right ]
\label{eq2_0}
\end{equation}

An energy-dissipation function of the process ${\mathcal E}_d[p^+,p^-](t)$
is a bilinear functional of the partial probability densities
$p^+(x,t)$, $p^-(x,t)$,
\begin{equation}
{\mathcal E}_d(t)={\mathcal E}_d[p^+,p^-](t) = \sum_{\alpha,\beta=\pm} C_{\alpha,\beta}
\int p^{\alpha}(x,t) \, p^{\beta}(x,t) \, d x
\label{eq7_1}
\end{equation}
where $C_{\alpha,\beta}$ are non-negative constants,
such that along the evolution of the process
\begin{equation}
\frac{d {\mathcal E}_d(t)}{d t} \leq 0
\label{eq7_2}
\end{equation}
where, for notational simplicity, the explicit dependence on the
partial waves has been omitted.

Consider the unbounded propagation in  $x \in (-\infty,\infty)$.
In order to obtain an expression for ${\mathcal E}_d(t)$
take the system of eqs. (\ref{eq2_0}), multiply the
evolution equation for $p^+(x,t)$ by $p^+(x,t)$, and that for $p^-(x,t)$ 
by $p^-(x,t)$, sum them
together, and integrate over $x$,
\begin{eqnarray}
\frac{1}{2}  \frac{d }{d t} \int_{-\infty}^\infty
 \left [ (p^+)^2 + (p^-)^2 \right ] \, dx
 & = & - \frac{b}{2} \int_{-\infty}^\infty \partial_x \left [ (p^+)^2 - (p^-)^2
\right ] \, dx \nonumber \\
& - & \lambda \,  \int_{-\infty}^\infty (p^+-p^-)^2 \, d x
\label{eq7_3} 
\end{eqnarray}
where the regularity conditions at infinity have been enforced.
This means that an energy-dissipation function can be defined
as
\begin{equation}
{\mathcal E}_d(t) = \frac{1}{2} \int_{-\infty}^\infty \left [
(p^+)^2 + (p^-)^2 \right ] \, d x
\label{eq7_4}
\end{equation}
and eq. (\ref{eq7_3}) implies that
\begin{equation}
\frac{d {\mathcal E}_d(t)}{d t} = - \frac{\lambda}{b^2} \, || J_d ||_{L^2}^2(t)
\label{eq7_5}
\end{equation}
where $||f||_{L^2}^2(t)$, for  a real-valued
 square summable function $f(x,t)$,
is the square of its $L^2$-norm $||f||_{L^2}^2(t)= \int_{-\infty}^\infty
f^2(x,t) \, d x$.
Since $p^{\pm}= (p\pm J_d)/2$, ${\mathcal E}_d(t)$
can be expressed as
\begin{equation}
{\mathcal E}_d = \frac{1}{4} \int_{-\infty}^\infty 
\left ( p^2 + \frac{J_d^2}{b^2} \right ) \, d x
\label{eq7_6}
\end{equation}
i.e., it is a quadratic functional of both the overall
probability density function $p(x,t)$
and of its diffusive flux $J_d(x,t)$.
In the Kac limit, $\lambda/b^2= 1/2 D_{\rm eff}$, $J_d(x,t)=- D_{\rm eff}
 \partial_x p(x,t)$,
and eq. (\ref{eq7_5}) reduces to the classical
Fickian dissipation relation $\partial_t
 ||p||_{L^2}^2 = - 2 \, D_{\rm eff} \, ||\partial_x p||_{L^2}^2$. The remarkable property of eq. (\ref{eq7_5}) is that
the energy dissipation function  depends also on the
flux, which is the fundamental starting point in the theory of extended irreversible thermodynamics.

Next consider, instead of unbounded propagation, a closed interval $x \in
[0,L]$,  where zero-flux conditions applies at the boundaries $x=0,L$.
These conditions
 correspond to the
reflection conditions $p^+|_{x=0}=p^-|_{x=0}$, $p^-|_{x=L}=p^+|_{x=L}$
for the  partial probability waves. 
Eq. (\ref{eq7_3}) holds also in this case, substituting the integration
extremes $x=-\infty$ and $x=\infty$ with $x=0$ and $x=L$, respectively.
Observe that the reflection conditions make the divergence integral
appearing in eq. (\ref{eq7_3}) identically
equal to zero, so that  eqs. (\ref{eq7_5})-(\ref{eq7_6})
hold true also in this case, substituting the integration 
extremes  $-\infty$ and $\infty$
with $0$ and $L$, respectively.

Next, consider the entropy function. As a candidate for
 entropy consider
the Boltzmann-Shannon entropy $S_{BS}(t)$ defined with respect to
the partial  probability 
waves. In the case of unbounded propagation it reads
\begin{equation}
S_{BS}(t)= - \int_{-\infty}^\infty \left [
p^+ \, \log p^+ + p^- \, \log p^- \right ]  d x
\label{eq7_7}
\end{equation}
Enforcing the balance equations for the partial waves, one obtains
\begin{eqnarray}
\frac{d S_{BS}(t)}{d t} & = & - b \, \int_{-\infty}^\infty
\partial_x \left [ p^+ \, \log p^+ - p^- \, \log p^- - p^+ + p^- \right ]
 d x  \nonumber \\
& = & \lambda \, \int_{-\infty}^\infty (p^+- p^-) \, \log \left (\frac{p^+}{p^-}
\right ) \, d x
\label{eq7_8}
\end{eqnarray}
The divergence integral vanishes because of the regularity conditions at infinity, so that
\begin{equation}
\frac{d S_{BS}(t)}{d t} = \lambda \, \int_{-\infty}^\infty (p^+- p^-) \, \log \left (\frac{p^+}{p^-}
\right ) \, d x \geq 0
\label{eq7_9}
\end{equation}
since the function $g(x,y)=(x-y) \log(x/y)$ is non negative for $x,y \geq 0$.
An analogous result holds in a closed bounded system represented
by the interval $[0,L]$, in the presence of reflecting
boundary conditions for the partial waves, substituting the
integral from $-\infty$ to $\infty$ with an integral from $0$ to $L$

The first  analytical results for the entropy function in the presence
of a hyperbolic Cattaneo transport  model have been
derived by Camacho and Jou \cite{entropy1}, exhibiting a quadratic
dependence on the probability flux, and reducing under equilibrium
condition to the Boltzmann $H$-function. This result has been
generalized by Vlad and Ross for telegrapher-type master
equations \cite{entropy2}. A recent survey on the
entropy principle related with extended thermodynamic formulations
can be found in \cite{entropy3}.

\subsection{GPK processes}
\label{sec_2_2}

The analysis of dissipation functions and entropies in
GPK processes is essentially related to  the underlying Markov-chain
structure of the finite $N$-state  Poisson process generating stochasticity
in the system. The key role is played by the spectral properties
of the transition matrix ${\bf A}$ which, in the general
setting, is simply an irreducible
 left-stochastic matrix $A_{\alpha,\beta} \geq 0$,
$\sum_{\gamma=1}^N A_{\gamma,\alpha}=1$, $\alpha=1,\dots,N$.
The spectral properties of ${\bf A}$ that are relevant in the
remainder are: (i) the spectral radius of ${\bf A}$ is
1 \cite{seneta}, i.e., all the eigenvalues $\mu_\alpha$, $\alpha=1,\dots,N$
of ${\bf A}$, i.e., $\sum_{\beta=1}^N A_{\alpha, \beta}  \, c_\beta = \mu_\alpha
\, c_\alpha$,
are such that $|\mu_\alpha|\leq 1$; (ii) the dominant Frobenius eigenvalue
is $\mu=1$, corresponding to a uniform left eigenvector (all the
entries are equal); (iii)    for $\alpha=2,\dots,N$, $\mu_\alpha < 1$.

From part I we know that the statistical description of
a GPK process defined by $N$ distinct constant stochastic
velocity vectors ${\bf b}_\alpha$, $\alpha=1,\dots, N$,
in the presence of a deterministic velocity field ${\bf v}({\bf x})$,
involves $N$ partial probability density functions $p_\alpha({\bf x},t)$,
$\alpha=1,\dots,N$
satisfying the hyperbolic system of equations
\begin{eqnarray}
\hspace{-1.0cm} \partial_t p_{\alpha}({\bf x}, t)
 =   - \nabla \cdot \left [
({\bf v}({\bf x})+ {\bf b}_\alpha) \,
p_{\alpha}({\bf x}, t) \right ]
 - \lambda_\alpha p_{\alpha}({\bf x}, t)
+ \sum_{\gamma=1}^N \lambda_\gamma \, A_{\alpha,\gamma}
\, p_{\gamma}({\bf x}, t) 
\label{eq2_1}
\\
\nonumber
\end{eqnarray}

Throughout this paragraph, we assume for simplicity
that the deterministic velocity field ${\bf v}({\bf x})$ is
solenoidal, i.e., $\nabla \cdot {\bf v}({\bf x})=0$.
 This
condition, with some further technical efforts, could be  
removed, at least for some classes of  potential and mixed flows. The
generalization to generic irrotational velocity fields (potential flows)
 is left open, and is not as simple as it may seem, for
technical reasons that are briefly addressed in Section \ref{sec_4}.

The analysis of energy dissipation and entropy functions for GPK
processes is developed gradually by considering classes
of processes of increasing structural complexity, defined
by the symmetry properties of ${\mathcal B}_N$, ${\bf A}$ and $\boldsymbol{\Lambda}$ (see part I).

To begin with, consider the simplest case of a  transitionally
symmetric GPK process possessing
a uniform transition rate vector ${\boldsymbol \Lambda}=
(\lambda,\dots,\lambda)$. In this, case the transition probability
matrix ${\bf A}$
is also symmetric. For this class of processes, an energy dissipation
function is given by
\begin{equation}
{\mathcal E}_d[\{p_\alpha \}_{\alpha=1}^N](t)= \frac{1}{2}
\sum_{\alpha=1}^N \int_{{\mathbb R}^n} p_\alpha^2({\bf x},t) \, d {\bf x}
\label{eq7_10}
\end{equation}
To prove this, multiply each balance equation for the
corresponding partial wave $p_\alpha({\bf x},t)$, sum over the
states $\alpha$ and integrated with respect to ${\bf x}$ to
obtain
\begin{eqnarray}
\frac{d {\mathcal E}_d(t)}{d t} &= & - \sum_{\alpha=1}^N
\int_{{\mathbb R}^n} p_\alpha
\, \nabla \cdot \left ( {\bf v} \, p_\alpha \right )
\, d {\bf x} - \sum_{\alpha=1}^N \int_{{\mathbb R}^n}
 p_\alpha \, \nabla \cdot
\left ( {\bf b}_\alpha \, p_\alpha \right )  \, d {\bf x} \nonumber \\
 & -  & \lambda \, \sum_{\alpha=1}^N \int_{{\mathbb R}^n}
 p_\alpha^2 \, d {\bf x} +
\lambda \sum_{\alpha,\gamma=1}^N \int_{{\mathbb R}^n}
 p_\alpha \, A_{\alpha,\gamma} \, p_\gamma
\, d {\bf x}
\label{eq7_11}
\end{eqnarray}
The first two integrals vanish as they can be expressed in
a divergence form, $p_\alpha \, \nabla \cdot \left ({\bf v} \, p_\alpha
\right )= \nabla \cdot \left ({\bf v} \, p_\alpha^2/2 \right )$,
$p_\alpha \, \nabla \cdot \left ({\bf b}_\alpha \, p_\alpha
\right )= \nabla \cdot \left ({\bf b}_\alpha \, p_\alpha^2/2 \right )$,
and regularity conditions at infinity apply.

Indicating with $({\bf f},{\bf g})_{L^2_N}$ the scalar product for
$N$-dimensional real-valued 
 square summable functions ${\bf f}({\bf x})=(f_1({\bf x}),\dots,f_N({\bf x}))$,
${\bf g}({\bf x})=(g_1({\bf x}),\dots,g_N({\bf x}))$ of ${\mathbb R}^n$,
\begin{equation}
({\bf f},{\bf g})_{L_N^2}= \sum_{\alpha=1}^N \int_{{\mathbb R}^n}
 f_\alpha({\bf x}) \,
g_\alpha({\bf x}) \, d {\bf x}
\label{eq7_12}
\end{equation}
eq. (\ref{eq7_11}) can be expressed as
\begin{equation}
\frac{1}{\lambda} \, \frac{d {\mathcal E}_d(t)}{d t}=
- ({\bf p},{\bf p} )_{L_N^2} + ({\bf A} \, {\bf p},{\bf p})_{L^2_N}
\label{eq7_13}
\end{equation}
where ${\bf p}({\bf x},t)=(p_1({\bf x},t),\dots,p_N({\bf x},t))$
is the vector of the partial probability waves.
Since the maximum (Frobenius) eigenvalue of ${\bf A}$ equals $1$, and all
the other eigenvalues lie within the unit circle
and   possess real parts less than $1$, it follows that
\begin{equation}
|({\bf A} \, {\bf p},{\bf p})_{L^2_N} | \leq ({\bf p},{\bf p} )_{L_N^2}
\label{eq7_14}
\end{equation}
and consequently, eq. (\ref{eq7_13}) provides the
inequality
\begin{equation}
\frac{d {\mathcal E}_d(t)}{d t} \leq 0
\label{eq7_15}
\end{equation}
As regards the entropy function, one can 
consider the Boltzmann-Shannon expression defined
starting from the partial probability waves characterizing the
process
\begin{equation}
S_{BS}(t)= - \sum_{\alpha=1}^N \int_{{\mathbb R}^n} p_\alpha({\bf x},t) \,
\log p_\alpha({\bf x},t) \, d {\bf x}
\label{eq7_16}
\end{equation}
Enforcing the conservation property
$\sum_{\alpha=1}^N \int_{{\mathbb R}^n} p_\alpha({\bf x},t) d {\bf x}=1$,
and simplifying the resulting
expression as regards the divergence terms that are  vanishing
because of the regularity at infinity, one finally obtains
\begin{eqnarray}
\frac{d S_{BS}(t)}{d t} &= & \lambda \left [
\sum_{\alpha=1}^N \int_{{\mathbb R}^n} p_\alpha \, \log p_\alpha \, d {\bf x}
- \sum_{\alpha,\gamma=1}^N \int_{{\mathbb R}^n}
 \log p_\alpha \, A_{\alpha,\gamma}
\, p_{\gamma} \, d {\bf x}  \right ] \nonumber \\
& = & \lambda \left [ \left ({\bf p},\log {\bf p} \right )_{L^2_N}
- \left ( {\bf A} \, {\bf p} , \log {\bf p} \right )_{L_N^2} \right ]
\label{eq7_17}
\end{eqnarray}
where we have set $\log {\bf p}=(\log p_1,\dots,\log p_N)$.
The term at the right-hand side of eq. (\ref{eq7_17})
equals $\lambda$ times the integral over ${\bf x}$ of  a 
function $g_S({\bf p})$, i.e., $dS_{BS}(t)/dt= \lambda \, \int_{{\mathbb R}^n}
g_S({\bf p}({\bf x},t)) \, d {\bf x}$, given by
\begin{eqnarray}
g_S({\bf p}) & = & \frac{1}{2} \sum_{\alpha,\gamma=1}^N
A_{\alpha,\gamma} \, (p_\alpha-p_\gamma) \, \log \left ( \frac{p_\alpha}{p_\gamma} \right ) \nonumber \\
& = & \frac{1}{2} \left [
\sum_{\alpha,\gamma=1}^N A_{\alpha,\gamma} \, p_\alpha \, \log p_{\alpha}
-\sum_{\alpha,\gamma=1}^N A_{\alpha,\gamma} \, p_\gamma \, \log p_{\alpha}
\right . \nonumber \\
&- & \left . \sum_{\alpha,\gamma=1}^N A_{\alpha,\gamma} \, p_\alpha
\, \log p_\gamma + \sum_{\alpha,\gamma=1}^N A_{\alpha,\gamma}
\, p_\gamma \log p_\gamma \right ] \nonumber \\
& = &   \sum_{\alpha=1}^N p_\alpha \, \log p_\alpha - \sum_{\alpha,\gamma=1}^N
\log p_\alpha \, A_{\alpha,\gamma} \, p_\gamma
\label{eq7_18}
\end{eqnarray}
where the symmetry of $A_{\alpha,\gamma}$ and its left stochasticity
have been enforced. Since $A_{\alpha,\gamma} \geq 0$,
and each factor  $(p_\alpha-p_\gamma) \, \log(p_\alpha/p_\gamma)$
is greater than or at most equal to zero for $p_\alpha({\bf x},t),
p_\gamma({\bf x},t) \geq 0$, it follows that
\begin{equation}
\frac{d S_{BS}(t)}{d t} \geq 0
\label{eq7_19}
\end{equation}
Next, consider the general case in the presence of an arbitrary 
distribution of transition rate $\lambda_\alpha$, $\alpha=1,\dots,N$.
As regards the energy-dissipation function, it is convenient to
introduce the auxiliary functions $u_\alpha({\bf x},t)$ defined
as
\begin{equation}
u_\alpha({\bf x},t) = \lambda_\alpha \, p_\alpha({\bf x},t)
\, , \qquad \alpha=1,\dots,N
\label{eq7_20}
\end{equation}
 In terms of  the $u_\alpha$'s, the balance equations
(\ref{eq2_1}) become
\begin{eqnarray}
\lambda_\alpha^{-1} \, \partial_t u_\alpha  =  
-\lambda_\alpha^{-1} \nabla \cdot
\left ( {\bf v} \, u_\alpha \right )
- \lambda_\alpha^{-1} \, \nabla \cdot \left ( {\bf b}_\alpha \, u_\alpha
\right ) 
 - u_\alpha + \sum_{\gamma=1}^N A_{\alpha,\gamma} \, u_\gamma
\label{eq7_21}
\end{eqnarray}
$\alpha=1,\dots,N$.
It is natural to introduce the following energy dissipation
function
\begin{equation}
{\mathcal E}_d[\{p_\alpha \}_{\alpha=1}^N](t) = 
\sum_{\alpha=1}^N \frac{1}{2 \, \lambda_\alpha} \int_{{\mathbb R}^n} u_\alpha^2({\bf x},t)
\, d {\bf x}= \sum_{\alpha=1}^N  \frac{\lambda_\alpha}{2}
\int_{{\mathbb R}^n} p_\alpha^2({\bf x},t) \, d {\bf x}
\label{eq7_22}
\end{equation}
Performing  the  same algebra as in the previous case and setting
${\bf u}=(u_1,\dots,u_N)$,  one
obtains
\begin{equation}
\frac{d {\mathcal E}_d(t)}{d t}= -\left ({\bf u},{\bf u} \right )_{L^2_N}
+ \left ( {\bf A} \, {\bf u},{\bf u} \right )_{L^2_N} \leq 0
\label{eq7_23}
\end{equation}
that follows from the fact that ${\bf A}$ is  left-stochastic.
Observe that no assumptions has been made on the symmetry
of the transition matrix $K_{\alpha,\gamma}=\lambda_\gamma \, A_{\alpha,\gamma}$
so that  eqs. (\ref{eq7_22}) apply both for transitionally symmetric
and non-symmetric GPK processes.

Next, consider  the  entropy.
As regards the entropy function, the local detailed balance 
defining transitionally symmetric GPK processes counts.
To begin with, consider transitionally symmetric GPK processes,
characterized by the property that
the transition matrix 
${\bf K}= {\bf A} \, {\boldsymbol \Lambda}$ is symmetric, i.e.,
\begin{equation}
K_{\alpha,\gamma}= \lambda_\gamma \, A_{\alpha,\gamma}=
\lambda_\alpha \, A_{\gamma,\alpha}=K_{\gamma,\alpha}
\, , \qquad \alpha,\gamma=1,\dots,N
\label{eq7_24}
\end{equation}
with the property that $K_{\alpha,\gamma} \geq 0$ and
\begin{equation}
\sum_{\gamma=1}^N K_{\gamma,\alpha}= \lambda_\alpha
\label{eq7_25}
\end{equation}
For transitionally symmetric GPK processes the expression
(\ref{eq7_16}) for the Boltzmann-Shannon entropy is
still a valid candidate as the entropy function of the process.
Enforcing the properties (\ref{eq7_24})-(\ref{eq7_25})
of the transition matrix ${\bf K}$, 
the time derivative of the Boltzmann-Shannon
entropy becomes
\begin{equation}
\hspace{-1.0cm}
\frac{d S_{BS}(t)}{d t} = \int_{{\mathbb R}^n} \sum_{\alpha,\gamma=1}^N
K_{\alpha,\gamma} \left [ p_\alpha({\bf x},t)  - p_\gamma({\bf x},t) \right ]
\, \log p_\alpha({\bf x},t) 
 ] \, d {\bf x}
\label{eq7_26}
\end{equation}
The latter expression can be written in terms of
a entropy-rate density $r_S({\bf p}({\bf x},t))$, i.e., as
 $dS_{BS}(t)/dt=  \int_{{\mathbb R}^n}
r_S({\bf p}({\bf x},t)) \, d {\bf x}$,
 given by
\begin{equation}
r_S({\bf p}) = \frac{1}{2} \sum_{\alpha,\gamma=1}^N K_{\alpha,\gamma}
\, (p_\alpha-p_\gamma ) \, \log \left ( \frac{p_\alpha}{p_\gamma} \right )
\label{eq7_27}
\end{equation}
which,  by definition, is greater than or at most equal to zero for
any $p_{\alpha}({\bf x},t) \geq 0$, $\alpha=1,\dots,N$.

There is another situation of physical interest (see  further paragraph
\ref{sec_2_4}), namely when the transition probability
matrix $A_{\alpha,\gamma}$ is symmetric, but the transition
rates $\lambda_\alpha$, $\alpha=1,\dots,N$ 
are arbitrary positive constants. The resulting GPK process
is therefore transitionally non-symmetric.
In this case, one can define a modified Boltzmann-Shannon
entropy $\widehat{S}_{BS}(t)$  as
\begin{equation}
\hspace{-2.0cm} 
\widehat{S}_{BS}(t)= - \sum_{\alpha=1}^N \frac{1}{\lambda_\alpha}
\int_{{\mathbb R}^n} u_\alpha({\bf x},t) \, \log u_\alpha({\bf x},t)
\, d{\bf x}= - \sum_{\alpha=1}^N \int_{{\mathbb R}^n} p_\alpha({\bf x},t)
\, \log  \left [ \lambda_\alpha \, p_\alpha({\bf x},t) \right ]  \, d {\bf x}
\label{eq7_28}
\end{equation}
where $u_\alpha({\bf x},t)$ are defined by
eq. (\ref{eq7_20}). From the evolution equations
(\ref{eq7_21}) it follows after some algebra  that
\begin{equation}
\frac{d \widehat{S}_{BS}(t)}{ d t}= \left ({\bf u}, \log {\bf u} \right 
)_{L_N^2} - \left ({\bf A} {\bf u}, \log {\bf u} \right 
)_{L_N^2}
\label{eq7_29}
\end{equation}
where we have used the notation ${\bf u}=(u_1,\dots,u_N)$,
$\log {\bf u}=(\log u_1,\dots,\log u_N)$.
Eq. (\ref{eq7_29}) is formally analogous to a previously treated 
case, see eq. (\ref{eq7_17}), so that
\begin{equation}
\frac{d \widehat{S}_{BS}(t)}{ d t} = \int_{{\mathbb R}^n} 
r_S({\bf }u({\bf x},t))
\, d {\bf x}
\label{eq7_30}
\end{equation}
where
\begin{equation}
r_S({\bf u})= \frac{1}{2} \sum_{\alpha,\gamma=1}^N A_{\alpha,\gamma}
\, (u_\alpha-u_\gamma) \, \log \left ( \frac{u_\alpha}{u_\gamma} \right )
\geq 0
\label{eq7_31}
\end{equation}

\subsection{A simple example}
\label{sec_2_3}

\begin{figure}[h]
\begin{center}
{\includegraphics[height=6cm]{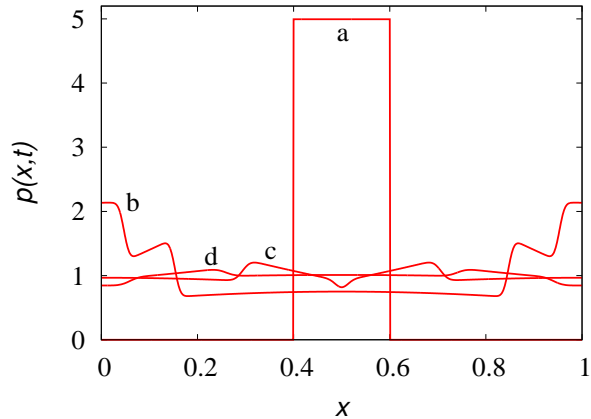}}
\end{center}
\caption{$p(x,t)$ vs $x$ at several time instant for $\lambda=10$.
Line (a) refers to the initial condition (\ref{eq7_2_1}),
line (b) to $t=0.1$, line (c) to $t=0.2$, line (d) to $t=0.3$.}
\label{Fig11}
\end{figure}

This paragraph highlights the dissipation properties
addressed in the previous paragraph through a simple example.
Consider the one-dimensional, purely stochastic ($v(x)=0$) Poisson-Kac
process  $d x(t) = b (-1)^{\chi(t)} \, dt$
 on the unit interval with reflective conditions
at the boundaries.
Keeping fixed $D_{\rm eff}=b^2/2 \lambda=1$, use the transition
rate $\lambda$ as a parameter.
As initial condition take
\begin{equation}
p^+(x,0)=p^-(x,0)= \left \{
\begin{array}{ccc}
1/4 d & & |x-1/2|\leq 0 \\
0 & & \mbox{otherwise}
\end{array}
\right .
\label{eq7_2_1}
\end{equation}
so that $\int_0^1 p(x,t) \, dx=1$ for $t \geq 0$.
The energy dissipation function introduced in the
previous paragraph can be normalized by
considering the auxiliary function
\begin{equation}
{\mathcal E}_d^*(t)= 2 {\mathcal E}_d(t) - \frac{1}{2}
\label{eq7_2_2}
\end{equation}
so that $\lim_{\rightarrow \infty} {\mathcal E}_d^*(t)=0$.
The Fickian counterpart of ${\mathcal E}^*(t)$ is
represented by
\begin{equation}
E^*(t)= \int_0^1 p^2(x,t) \, dx -1 = ||p-1||_{L^2}^2
\label{eq7_2_3}
\end{equation}
that corresponds to the square of the $L^2$-norm of the
overall probability density function normalized to zero mean.
Figure \ref{Fig11} depicts several concentration
profiles of the overall probability density function $p(x,t)$
for $\lambda=10$, sampled at time-intervals of $0.1$, just
to visualize the typical deviation from Brownian evolution
characterizing Poisson-Kac dynamics at short timescales.

\begin{figure}[h]
\begin{center}
{\includegraphics[height=8cm]{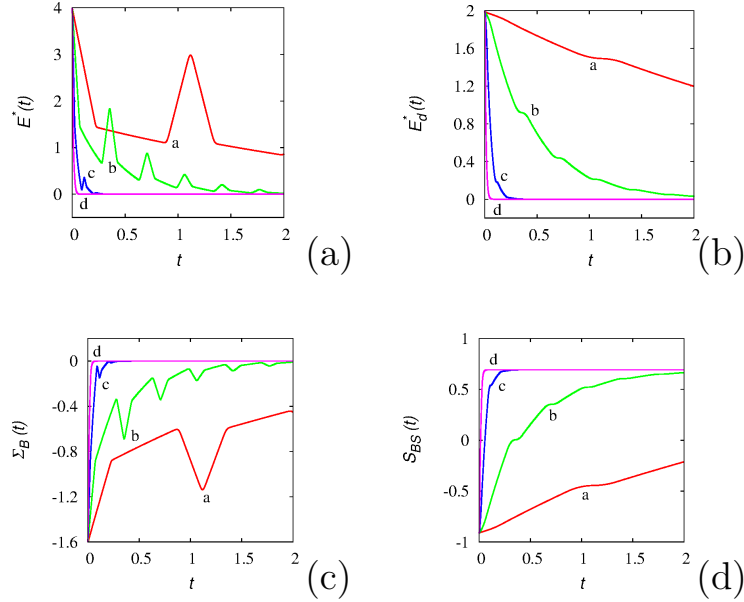}}
\end{center}
\caption{Review of the  time evolution
of dissipation and entropy functions
for the Poisson-Kac process considered in the main text at $D_{\rm eff}=1$.
Panel (a) refers to $E^*(t)$, panel (b) by to ${\mathcal E}_d^*(t)$,
panel (c) to the Boltzmannian entropy $\Sigma_B(t)$, panel (d)
to the Boltzmann-Shannon entropy $S_{BS}(t)$ defined
using the partial probability waves. Lines from (a) to (d)
in all the panels refer to $\lambda=0.1\,, 1,\, 10,\, 100$, respectively.} 
\label{Fig12}
\end{figure}

The comparison between the Fickian $E^*(t)$ and
the correct ${\mathcal E}_d^*(t)$  energy dissipation
functions is depicted in panels (a) and (b)
of figure \ref{Fig12}. While $E^*(t)$
exhibits an evident non-monotonic behavior as a function of $t$
 in the range of transition
rates $\lambda \in (0.1,10)$,
 that
becomes more pronounced as $\lambda$ decreases, the
function ${\mathcal E}_d^*(t)$ is monotonically non-increasing.
This example expresses pictorially the claim that an energy-dissipation
function which is a  quadratic
functional solely of the overall probability
density  cannot be compatible with Poisson-Kac dynamics,
and more generally with stochastic evolution possessing a finite
propagation velocity. Conversely, the
function ${\mathcal E}_d^*(t)$, that depends on the
whole system of partial waves  - in the present
 case  $p^+(x,t)$ and $p^-(x,t)$ -
provides a correct description of dissipation.
This example supports the fundamental ansatz of the theory
of extended thermodynamics that state and dissipation functions
in 
irreversible processes 
should depend also on the fluxes, as ${\mathcal E}_d^*(t)$ in the
present case \cite{ext1}.
At $\lambda=100$ (lines (d)) $E^*(t)$ and ${\mathcal E}_d^*(t)$ practically
coincide, and this corresponds to the Kac limit of the
process.

A specular behavior is displayed by the entropy function.
In this framework, the behavior of the
Boltzmann-Shannon entropy $S_{BS}(t)$ based on the
full structure of the partial probability waves should be contrasted with
the classical Boltzmannian entropy $\Sigma_B(t)$
\begin{equation}
\Sigma_B(t) = - \int_0^1 p(x,t) \, \log p(x,t) \, d x
\label{eq7_2_4}
\end{equation}
depending exclusively on the overall probability density
function $p(x,t)$. Panel (c) and (d) of figure
\ref{Fig12} show the comparison of these two entropy functions.
A similar analysis based on the Cattaneo equation has been performed 
by Jou et al. \cite{ext1}.
All the observations addressed for the energy dissipation functions
apply {\em verbatim} to $\Sigma_B(t)$ and $S_{BS}(t)$.
In the long-term limit $\Sigma_B(t) \rightarrow 0$,
while $S_{SB}(t) \rightarrow \log 2$, corresponding to
the complete homogenization amongst the partial waves.

\subsection{Relativistic transformation of entropy}
\label{sec_2_4}

At the end of paragraph \ref{sec_2_2} the expression for the entropy
of a GPK process possessing a symmetric transition probability
matrix and generic transition rates has been obtained, eq. (\ref{eq7_28}).
This case find application in relativistic analysis of stochastic
processes as outlined below.

Consider a one-dimensional, purely diffusive Poisson-Kac dynamics
in an inertial reference frame $\Sigma$, defined by the space-time coordinates
$(x,t)$, $d x(t)= b \, (-1)^{\chi(t)} \, dt$, in which the evolution
equations of the partial probability waves of the process $p^\pm(x,t)$
is expressed by eq. (\ref{eq2_0}). The frame $\Sigma$ can be referred
to as the {\em rest frame} of the process, since the process
is characterized by a vanishing effective velocity (corresponding
to the time-derivative of the first-order moment in the long-term
regime).

Let $\Sigma^\prime$ be another  inertial frame, defined
by the space-time coordinates $(x^\prime,t^\prime)$ moving
with respect to $\Sigma$ at constant relative velocity $v < c$,
where $c$ is the velocity of light {\em in vacuo}.
Enforcing the Lorentz boost connecting $(c \, t^\prime, x^\prime)$ to
$( c \,  t,x)$,
\begin{equation}
\left (
\begin{array}{c}
c \, t^\prime \\
x^\prime
\end{array}
\right )
= \gamma(v) \, \left (
\begin{array}{cc}
1 & -\beta \\
-\beta & 0
\end{array}
\right )
\, 
\left (
\begin{array}{c}
c \, t \\
x
\end{array}
\right )
\label{eq_m1}
\end{equation}
where $\gamma(v)=1/\sqrt{1-v^2/c^2}$ is the Lorentz factor and $\beta=v/c$,
the statistical description of the process in $\Sigma^\prime$
involves the partial probability density functions $p^{\pm,\prime}(x^\prime,t^\prime)$ that satisfy the balance equations \cite{giona_rel1,giona_rel2}
\begin{eqnarray}
\partial_{t^\prime} p^{+,\prime}(x^\prime,t^\prime) & 
= &  - b_+^\prime \, \partial_{x^\prime} p^{+,\prime}(x^\prime,t^\prime) 
- \lambda_+^\prime p^{+,\prime}(x^\prime,t^\prime)+\lambda_-^\prime p^{-,\prime}(x^\prime,t^\prime) \nonumber \\
\partial_{t^\prime} p^{-,\prime}(x^\prime,t^\prime) & 
= &  - b_-^\prime \, \partial_{x^\prime} p^{-,\prime}(x^\prime,t^\prime) 
+ \lambda_+^\prime p^{+,\prime}(x^\prime
,t^\prime)- \lambda_-^\prime p^{-,\prime}(x^\prime,t^\prime) 
\label{eq_m2}
\end{eqnarray}
where the velocities $b_\pm^\prime$ satisfy the usual relativistic
velocity transformation for $b$ and $-b$, respectively,
\begin{equation}
b_+^\prime = \frac{b-v}{1-b v/c^2} \, , \qquad
b_-^\prime = \frac{-b-v}{1+b v/c^2}
\label{eq_m3}
\end{equation}
while the transition rates $\lambda_\pm^\prime$ in $\Sigma^\prime$
are expressed by the relations (see \cite{giona_rel1,giona_rel2})
\begin{equation}
\lambda_+^\prime = \frac{\lambda}{\gamma(v)} \left ( 1- \frac{b \, v}{c^2}
\right )^{-1} \, ,
\qquad
\lambda_-^\prime = \frac{\lambda}{\gamma(v)} \left ( 1 + \frac{b \, v}{c^2}
\right )^{-1}
\label{eq_m4}
\end{equation}
The Lorentz boost does not change the transition probability matrix,
that in the present case is ${\bf A}^\prime={\bf A}=
\left ( \begin{array}{cc} 0 & 1 \\ 1 & 0 \end{array} \right )$,
but modifies the transition rates $\lambda_+^\prime$,
$\lambda_-^\prime$ in $\Sigma^\prime$. The transition rates 
 $\lambda_+$,
$\lambda_-$ coincide  at $v=0$
but, as the velocity $v$ increases, become
progressively more different from each other.
In $\Sigma^\prime$ the stochastic process considered
is still a GPK process with uneven transition rates
and symmetric transition probability matrix, as
addressed at the end of paragraph \ref{sec_2_2}.
Consequently, a suitable expression for the
entropy function in a generic frame is given
by eq. (\ref{eq7_28}), i.e.
\begin{equation}
\hspace{-1.5cm} \widehat{S}_{BS}(t)= - \int_{-\infty}^\infty
\left [ p^+(x,t) \, \log (\lambda_+ p^+(x,t)) + p^-(x,t) \, \log (\lambda_- p^-(x,t))  \right ] \, d x
\label{eq_m5}
\end{equation}
In $\Sigma$, $\lambda_+=\lambda_-=\lambda$ and
eq. (\ref{eq_m5}) returns
\begin{equation}
\hspace{-1.5cm}  \widehat{S}_{BS}(t)= -
\int_{-\infty}^\infty \left [ p^+(x,t) \, \log  p^+(x,t) +
 p^-(x,t) \, \log  p^-(x,t)  \right ] \, d x -\log \lambda
= S_{BS}(t)-\log \lambda 
\label{eq_m6}
\end{equation}
while in $\Sigma^\prime$, the entropy function becomes
\begin{equation}
\hspace{-1.5cm}  \widehat{S}_{BS}^\prime(t^\prime)= 
- \int_{-\infty}^\infty
\left [ p^{+,\prime}(x^\prime,t^\prime) \, \log (\lambda_+^\prime p^{+,\prime}(x^\prime,t^\prime)) + p^{-,\prime}(x^\prime,t^\prime) \, \log (\lambda_-^\prime p^{-,\prime}(x^\prime,t^\prime)) \right  ] \, d x^\prime
\label{eq_m7}
\end{equation}

\begin{figure}[h]
\begin{center}
{\includegraphics[height=7cm]{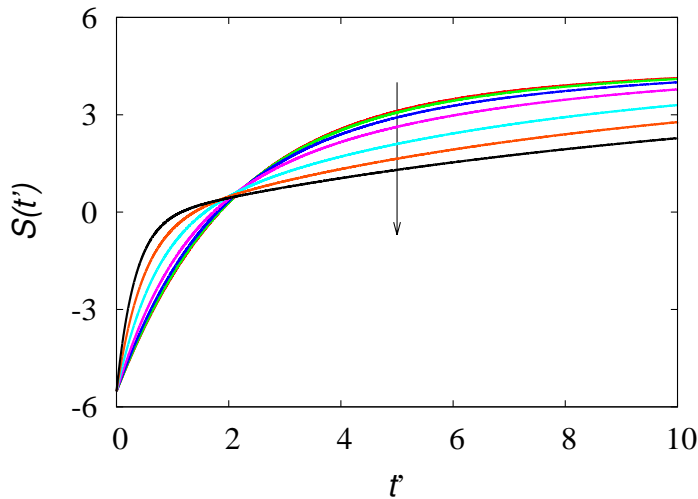}}
\end{center}
\caption{$S(t^\prime)=\widehat{S}_{BS}^\prime(t^\prime)$ vs $t^\prime$
for the Poisson-Kac process on the real line measured in
a reference system $\Sigma^\prime$ moving at constant relative
velocity $v$ with respect to the rest frame of the process.
The arrow indicates increasing values of $v=0,\,0.2,\,0.4,\,0.6,\,0.8,\,0.9,\,0.95$.}
\label{Fig_r1}
\end{figure}

Set $c=1$ a.u., and $b=c$, i.e., consider a stochastic perturbation the
characteristic velocity of which coincides with that of light,
as for electromagnetic fluctuations. Figure \ref{Fig_r1} depicts
the behavior of $\widehat{S}_{BS}^\prime(t^\prime)$ vs time $t^\prime$
for a Poisson-Kac process characterized in its rest frame
by $D_{\rm eff}=1$, i.e., $\lambda=b^2/2 D_{\rm eff}=1/2$.
The initial condition is symmetric and impulsive, namely
$p^+(x,0)=p^-(x,0)=\delta(x)/2$, centered at the origin. 
Apart from the monotonic behavior of $\widehat{S}_{BS}^\prime(t^\prime)$
with time $t^\prime$, it should be observed that the relativistic
transformation of the modified Boltzmann-Shannon
entropy cannot be easily expressed as a simple function of the
Lorentz factor $\gamma(v)$, as  occurs e.g. 
for   the tensor diffusivity \cite{giona_rel1}.

The extension to GPK processes is straightforward using the
relativistic transformation for  the partial probability densities
developed in \cite{giona_rel2} and the results at the end of paragraph
\ref{sec_2_2}.

\section{GPK processes and chaotic advection-diffusion problems}
\label{sec_3}
An interesting physical application of the GPK theory developed
above involves  tracer dynamics in chaotic flows
in the presence of stochastic perturbations.

Consider a two-dimensional problem defined
on the unit two-torus ${\mathcal T}^2=[0,1] \times [0,1]$, equipped
with periodic boundary conditions.
Let ${\bf v}({\bf x},t)$, ${\bf x}=(x,y)$ be a time-periodic
solenoidal velocity field, $\nabla \cdot {\bf v}=0$, 
and consider the GPK process
\begin{equation}
d {\bf x}(t)= {\bf v}({\bf x}(t),t) \, dt + \frac{1}{Pe} \, {\bf b}_{\chi_N(t)}
\, dt
\label{eq7_4_1}
\end{equation}
Equation (\ref{eq7_4_1}) represents the dimensionless kinematic
equations of motion of a passive tracer in an incompressible flow
subjected to stochastic (thermal) agitation expressed
by a finite $N$-state Poisson process acting
on a family of $N$ stochastic  velocity vectors.
The parameter $Pe$ is the P\'eclet number, representing
the ratio of the characteristic diffusion 
to the characteristic advection times.

Assume for the $N$-state finite Poisson process a constant
transition rate $\lambda$, and a transition probability
 matrix expressed by 
$A_{\alpha,\beta}=1/N$, $\alpha,\beta=1,\dots,N$. For the stochastic velocity vectors ${\bf b}_\alpha$
choose the family given by eq. (78) in part I, so that
$D_{\rm nom}=1$.
As regards the velocity field, consider a simple but widely
used model of Hamiltonian chaos, originating
from the standard map ${\bf x}^\prime = \boldsymbol{\Phi}({\bf x})$ \cite{sm1,sm2}, expressed by
\begin{equation}
\left \{
\begin{array}{lll}
x^\prime = x+\frac{\nu}{2 \pi} \, \sin(2 \pi y) & & \mbox{mod.} \; 1 \\
y^\prime = y + x^\prime & & \mbox{mod.} \; 1
\end{array}
\right .
\label{eq7_4_2bis}
\end{equation}
where $\nu>0$ is a real parameter. In a continuous
time setting, the standard map can be recovered
as the stroboscopic map associated with the time-periodic
incompressible flow possessing period $T=2$ obtained
from the periodic repetition of the flow protocol
\begin{equation}
{\bf v}({\bf x},t)= \left \{
\begin{array}{lll}
( \frac{\nu}{2 \pi} \, \sin(2 \pi y), 0) & & 0 \leq t < 1 \\
(0,  x) & & 1 \leq t < 2 
\end{array}
\right .
\label{eq7_4_2}
\end{equation}
and corresponding to the periodic 
switching of two shear 
flows
along the $x$- and $y$-coordinates, respectively,
 the first of which is sinusoidally modulated.
Observe that the second flow, is not continuous on the
torus, while the resulting stroboscopic map is $C^\infty$.

\begin{figure}[h]
\begin{center}
{\includegraphics[height=6cm]{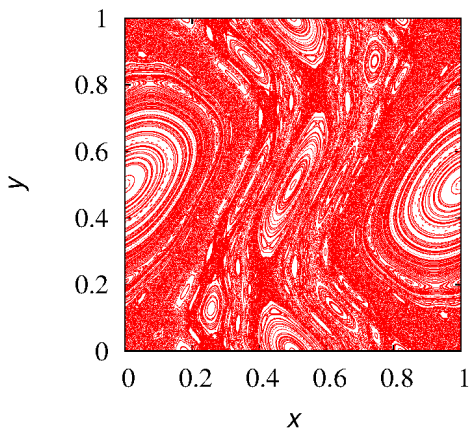}}
\end{center}
\caption{Poincar\'{e} map of the standard map at $\nu=1$.}
\label{Fig_x1}
\end{figure}

By varying the parameter $\nu$, the typical phenomenologies
of chaotic advection can be recovered from the
standard map. We consider the case $\nu=1$, the Poincar\'e map
of which (i.e., the stroboscopic map sampled at the period of
the flow protocol) is depicted in figure \ref{Fig_x1}, and is characterized
by the presence of invariant chaotic regions
possessing a maximum positive Lyapunov exponent,
intertwined with regular invariant islands of different sizes.

In the Kac limit, the statistical characterization
of eq. (\ref{eq7_4_1}) converges to the
solution of a classical parabolic advection-diffusion equation for the
overall probability density function $p({\bf x},t)$,
\begin{equation}
\partial_t p({\bf x},t) = - {\bf v}({\bf x},t) \cdot \nabla p({\bf x},t)
+ \frac{1}{Pe} \nabla^2 p({\bf x},t)
\label{eq7_4_3}
\end{equation}
Consider as an initial condition a completely
segregated initial profile of the partial probability waves,
namely
\begin{equation}
p_\alpha({\bf x},0) = \left \{
\begin{array}{lll}
2/N & & 0 \leq x < 1/2 \\
0 & & 1/2 \leq x < 1
\end{array}
\right .
\qquad
\alpha=1,\dots,N
\label{eq7_4_4}
\end{equation}
and let $E^*(t)$ be the normalized $L^2$-norm of $p({\bf x},t)$
\begin{equation}
E^*(t)= \frac{||p({\bf x},t)-1||_{L^2}}{||p({\bf x},0)-1||_{L^2}}
\label{eq7_4_5}
\end{equation}
so that $E^*(0)=1$ and  $\lim_{t \rightarrow \infty} E^*(t)=0$.

\begin{figure}[h]
\begin{center}
{\includegraphics[height=6cm]{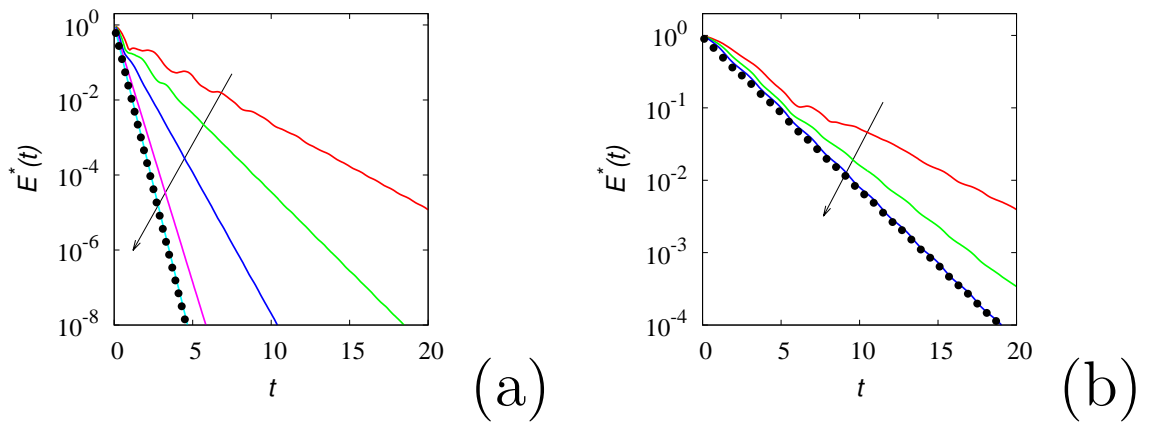}}
\end{center}
\caption{Normalized square $L^2$-norm $E^*(t)$ vs time $t$ for the
GPK flow  associated with the standard map ($N=4$) at $Pe=10^1$ (panel a), and $Pe=10^2$ (panel b).
Symbols ($\bullet$) represent the solution of the corresponding
parabolic 
advection-diffusion equation (\ref{eq7_4_3}). 
The  arrows indicate
increasing values of $\lambda$.
Panel (a):
$\lambda= 0.5, \, 1,\, 2,\,4,\,10,\,40$.
Panel (b):  $\lambda= 0.5, \, 1,\, 2$.}
\label{Fig13}
\end{figure}

Figure \ref{Fig13}  depicts the evolution of $E^*(t)$ 
and  at two
different values of the P\'eclet number:
$Pe=10^1$ (panel a) and $Pe=10^2$ (panel b).
Numerical simulations have been performed by expanding
$p_\alpha({\bf x},t)$ in truncated Fourier series,
$p_\alpha({\bf x},t)=\sum_{h,k=-N_c}^{N_c} P_{\alpha,h,k} \, e^{i 2 \pi
(h x+ k y)}$,  solving the corresponding system of linear
differential equations for the Fourier coefficients $P_{\alpha,h,k}$
with an explicit 4-th order Runge-Kutta solver.
For the range of $Pe$ values considered $Pe \leq 10^2$, we choose $N_c=50$, which is
fully sufficient for an accurate description of the dynamics,
apart from the very early stages of the process.
These graphs refer to a GPK process with $N=4$ using the
transition rate $\lambda$ as parameter.
The case $Pe=10^1$ (panel a) is indicative of the typical
relaxation properties of GPK systems: the normalized
$L^2$-norm $E^*(t)$ decays asymptotically in an exponential
way  $E^*(t) \sim e^{-\mu(\lambda) \, t}$, but for
moderate values of $\lambda$, the decay exponent $\mu(\lambda)$
is a function of the transition rate $\lambda$ and is
 smaller that the limit value $\mu_\infty= \lim_{\lambda \rightarrow \infty}
\mu(\lambda)$.

\begin{figure}[h]
\begin{center}
{\includegraphics[height=7.5cm]{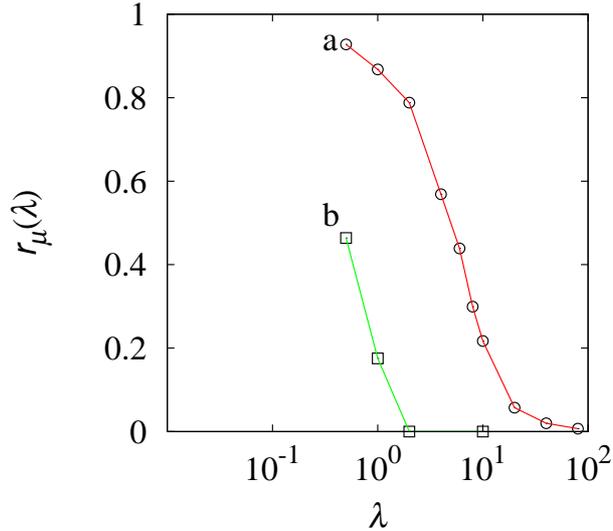}}
\end{center}
\caption{Ratio $r_\mu(\lambda)=[\mu_\infty-\mu(\lambda)]/\mu_\infty$ vs
$\lambda$ for the standard-map flow considered in the main text.
Line (a) and ($\circ$) refers to $Pe=10^1$, line (b) and ($\square$)
to $Pe=10^2$.}
\label{Fig_x2}
\end{figure}

As $\lambda$ increases,  the Kac-limit  property dictates that
$\mu(\lambda)$ converges towards the decay exponent $\Lambda(Pe)$ of
the parabolic advection-diffusion model (\ref{eq7_4_3}) for 
the same value of the P\'eclet number, i.e. $\mu_\infty=\Lambda(Pe)$.
The convergence of $\mu(\lambda)$ towards  $\mu_\infty$
is depicted in figure \ref{Fig_x2},
plotting the ratio $r_\mu(\lambda)=[\mu_\infty-\mu(\lambda)]/\mu_\infty$ 
vs $\lambda$ for the two
P\'eclet values considered. 
At $Pe=10^1$, the Kac convergence is achieved approximately
for $\lambda \geq 10^2$. At higher P\'eclet values, the
influence of $\lambda$ is less pronounced and 
the Kac-limit convergence is practically achieved
at smaller values of $\lambda$, e.g. $\lambda\geq 2$ for
$Pe=10^2$. 

\begin{figure}[h!]
\begin{center}
{\includegraphics[height=16cm]{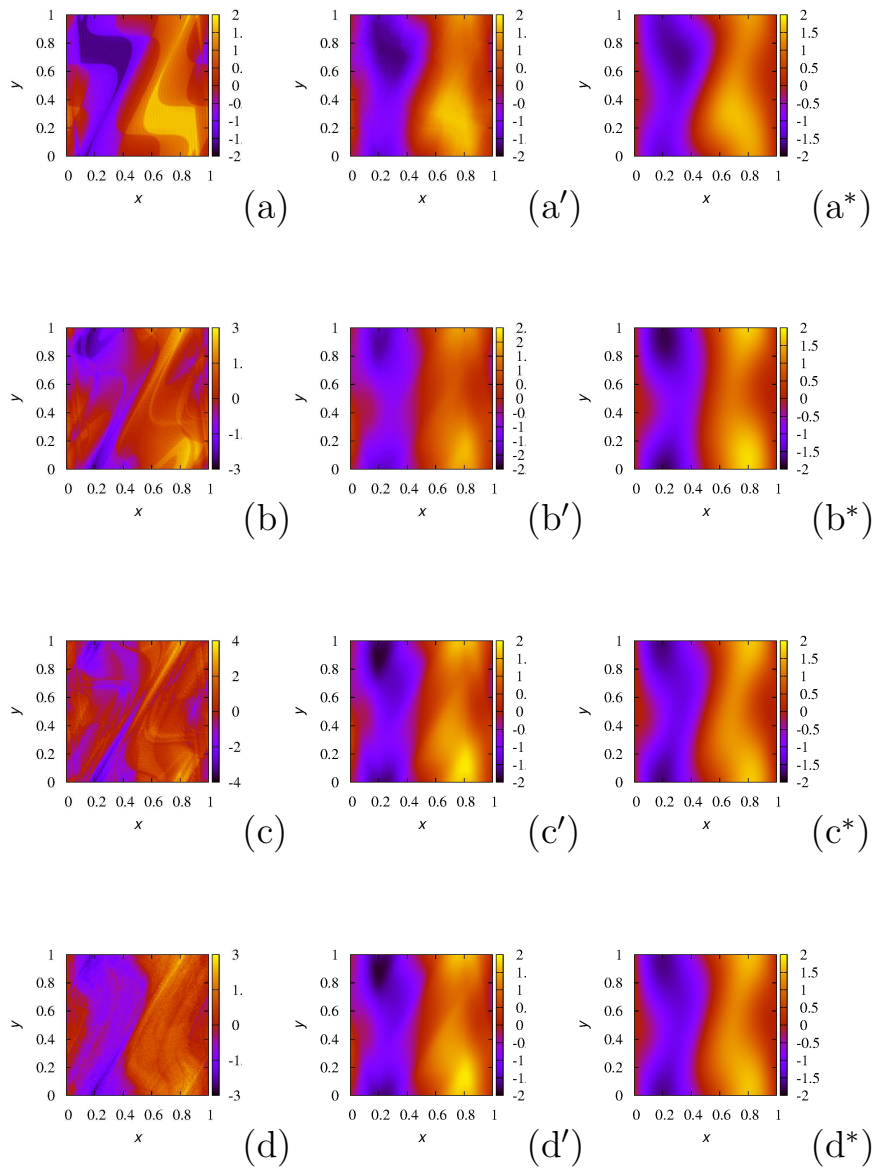}}
\end{center}
\caption{Concentration profiles of $p^*({\bf x},t)$
at  $Pe=10^2$, $N=4$ at $t=n T$.
Upper row $n=1$, second row $n=2$, third row $n=3$, and
lower row $n=10$.
Panels (a)-(d), first column, refer to $\lambda=1$
panels (a$^\prime$)-(d$^\prime$), second  column, to $\lambda=10$;
panels (a$^*$)-(d$^*$), third column, to the solution of the parabolic
 advection-diffusion
equation (\ref{eq7_4_3}).}
\label{Fig14}
\end{figure}

Figure \ref{Fig14} depicts a review of the early-time dynamics,
at  four time instants $t=n T$, $n=1,2,3,10$, at $Pe=10^2$,
 referred to the contour plots of
the rescaled overall probability density profiles
$p^*({\bf x},t)=C(p({\bf x},t)-1)$,
where $C$ is a normalization constant so that $p^*({\bf x},t)$
possess unit $L^2$-norm. Two  situations are considered:
far way from the Kac limit ($\lambda=1$), panels (a)-(d), and close
to the Kac limit ($\lambda=10$), panels (a$^\prime$)-(d$^\prime$), 
compared to the corresponding profiles obtained from the solution
of the parabolic advection-diffusion equation (\ref{eq7_4_3}),
panel (a$^*$),(d$^*$). The probability density profiles
at $\lambda=1$ still show a significant effect
of the hyperbolic (undulatory) dynamics
characterizing the evolution of the partial probability waves, as the
density profiles display much sharper discontinuities with
respect to the smoother behavior displayed by the
solutions of the parabolic equation (\ref{eq7_4_3}).
The graph for $t=n T=20$ depicted in the last row correspond
to the profile in asymptotic conditions of 
 the second eigenfunction of the Floqu\'et operator
associated with the advection-diffusion dynamics, see
\cite{giona_ces}.

\begin{figure}[h!]
\begin{center}
{\includegraphics[height=12cm]{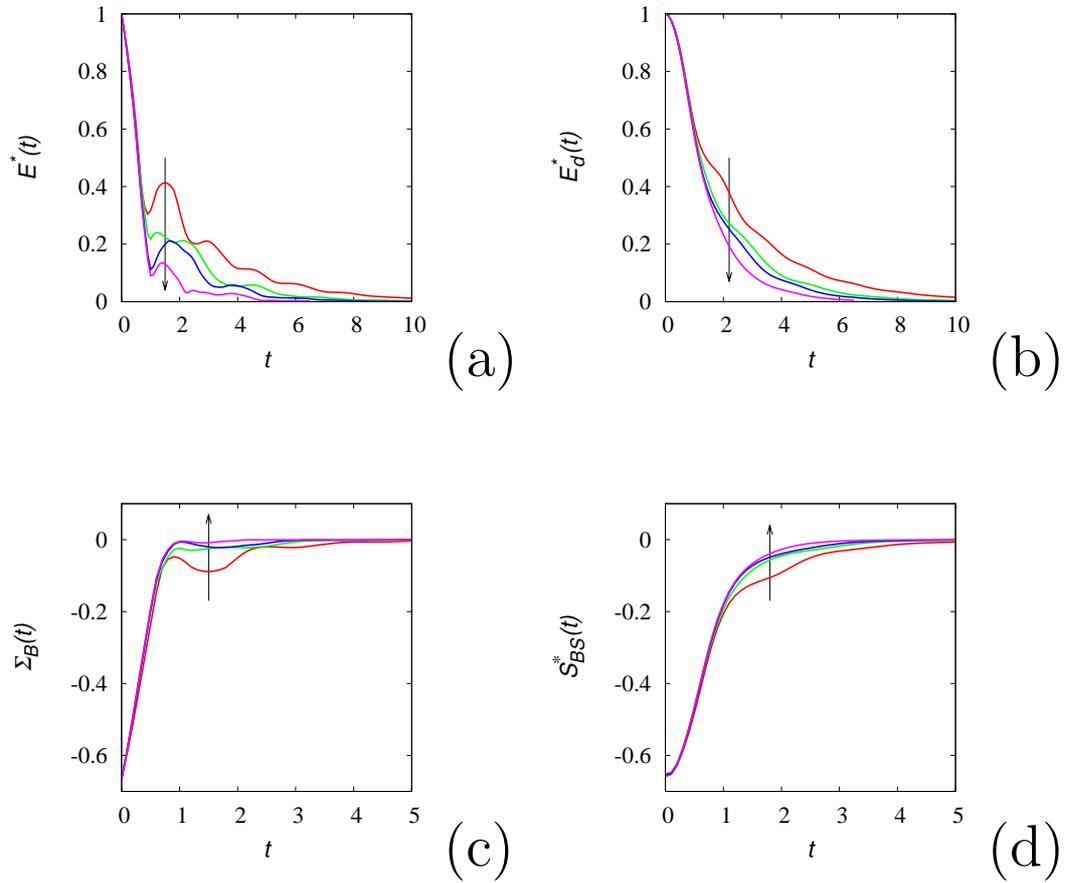}}
\end{center}
\caption{Review of the energy-dissipation functions/entropies
for the GPK processes associated
with the standard-map flow  at  $Pe=10^1$, $\lambda=1$.
The arrows in the four panels indicate increasing
values of $N=3,\,4,\,5,\,10$.
 Panels (a)-(b) depict energy-dissipation
functions: panel (a) refers to $E^*(t)$ vs $t$,
panel (b) to $E^*_d(t)$ vs $t$.
Panels (c)-(d) depict entropy functions:
panel (c) refers to the  Boltzmannian entropy $\Sigma_B(t)$ vs $t$,
panel (d) to the rescaled entropy $S_{BS}^*(t)$ based on the full
structure of the partial probability waves $\{p_\alpha({\bf x},t) \}_{\alpha=1}^N$.} 
\label{Fig15}
\end{figure}

Next, consider the energy dissipation functions and entropies.
A review of their behavior at $Pe=10^1$, $\lambda=1$ is
depicted in figure \ref{Fig15}, panels (a) to (d), using the
number $N$ of stochastic velocity vectors as parameter.
In these plots, $E^*_d(t)$ is the
normalized energy dissipation
function based on the partial probability waves
expressed as
\begin{equation}
E^*_d(t) = \frac{{\mathcal E}^*_d(t)}{{\mathcal E}_d^*(0)}
\qquad  {\mathcal E}_d^*(t) = \frac{1}{2} \sum_{\alpha=1}^N
\left | \left | p_\alpha({\bf x},t)-\frac{1}{N} \right | \right |_{L^2}
\label{eq7_4_6}
\end{equation}
while the normalized Boltzmann-Shannon entropy $S_{BS}^*(t)$
is the difference between the
Boltzmann-Shannon entropy $S_{BS}(t)$ and its limit value
$\log N$ for $t \rightarrow \infty$,
\begin{equation}
S_{BS}^*(t) = S_{BS}(t)- \log N 
\label{eq7_4_7}
\end{equation}
so that $\lim_{t \rightarrow \infty} S_{BS}^*(t) =0$
as for the Boltzmannian entropy $\Sigma_B(t)$. 

The comparisons of $E^*(t)$ and $E^*_d(t)$ (panels (a) and (b)) and
of $\Sigma_B(t)$ and  $S_{BS}^*(t)$ (panels (c) and (d))
indicate that the dissipation functionals $E^*(t)$ and $\Sigma_B(t)$
based exclusively on the overall probability density function $p({\bf x},t)$
display a non-monotonic/oscillatory behavior, while the corresponding quantities
$E_d^*(t)$ and $S_{BS}^*(t)$  based on the
full structure of the partial probability waves are
monotonic functions of time $t$. This is analogous to
the case of  the purely diffusive one-dimensional Poisson-Kac
model addressed in paragraph \ref{sec_2_3}.
There is however, a major conceptual difference between
the two problems, as regard the representation of the dissipation
functions.
In the one-dimensional problem treated in paragraph \ref{sec_2_3},
$p^+(x,t)$ adn $p^-(x,t)$ can  be expressed as a linear
combination of $p(x,t)$ and $J_d(x,t)$,
$p^{\pm}(x,t)= \left [ p(x,t) \pm J_d(x,t) \right ]/2$, 
 indicating that,
in the one-dimensional case in the presence of the two-state
process $(-1)^{\chi(t)}$, a correct energy dissipation function
and a consistent expression for the entropy can be always
expressed in terms of the overall probability density function $p(x,t)$
and of its diffusive flux $J_d(x,t)$.

This functional symmetry is broken in the
two-dimensional advection-diffusion problem considered
in this paragraph whenever $N \geq 4$. For $N \geq 4$,
the functional expressions for $E^*_d(t)$ and $S_{BS}^*(t)$
cannot be expressed exclusively in terms of $p({\bf x},t)$
and ${\bf J}_d({\bf x},t)$ as they depend on the
complete statistical structure of the GPK process, which is
accounted for by the system of partial probability waves $\{ p_{\alpha}({\bf x},t) \}_{\alpha=1}^N$.

This is a first, physically significant, case in which the 
concentration/flux paradigm characterizing the
classical theory of transport phenomena \cite{de_groot} results insufficient.
There is a further observation emerging from the
analysis of the data depicted in figure \ref{Fig15}.
The decay dynamics of the energy dissipation functions and entropies
depend on the number $N$ of stochastic velocity
vectors considered. However, as $N$ increases, a
limit behavior occurs, indicating that,
above a given  threshold $N^*$, the use of a higher number $N>N^*$
of states (velocity vectors) is practically immaterial. 

\section{Physical properties}
\label{sec_4}

In this Section, we address some physical observations
on the properties of GPK processes that can be of interest in several
 branches of physics.

\subsection{Stochastic field equations and Brownian-motion mollification}
\label{sec_4_1}

Starting from the works by Wong and Zakai \cite{wong_zakai1,wong_zakai2}, 
mollification (regularization)
of Wiener processes and Wiener-driven stochastic differential
equations has become an important field of stochastic
analysis.
In the original Wong-Zakai papers, Wiener processes
have been mollified using interpolation techniques obtaining
piecewise linear, and therefore almost everywhere (a.e.) 
smooth approximations of Brownian motion.
The so-called Wong-Zakai theorem derived by these authors
admits several important implications in stochastic
theory \cite{wong1,wong2,wong3} and the mollified version of a Langevin equation
is described statistically in a suitable limit by
the Stratonovich Fokker-Planck equation associated with
the original Langevin model, using the Stratonovich recipe
for the stochastic integrals.

Poisson-Kac and GPK processes provide a
physically significant way of mollifying stochastic
dynamics, as the Poisson-Kac perturbations, admitting
a finite propagation velocity, evolves as a physical
field, and possess a.e. regular trajectories. 
This property is particularly important in all
the cases, the stochastic perturbation does not derive by a coarse-grained approximation of many uncorrelated disturbances, but admits itself
a fundamental physical nature, such as the fluctuating
component of  the electromagnetic field (including 
the zero-point field), which plays
a central role in quantum electrodynamics, in understanding
fundamental particle-field interactions, and  in general cosmology
\cite{milonni}.

In this framework, GPK processes are the natural candidates for
attempting a modeling of fundamental field fluctuations,
since their wave-like propagation intrinsically match the requirement of
special relativity as it regards the bounds on the propagation velocity.
It is rather straightforward to derive from Poisson-Kac
and GPK
processes a Wong-Zakai theorem connecting the Kac
limit to the Stratonovich Fokker-Planck equation.

Mollification of Brownian motion  can be of 
 wide mathematical physical interest in connection
with the analysis of Stochastic Partial Differential
Equation (SPDE), that recently  experienced significant
progresses due to the introduction
of new concept and  mathematical tools
such as that of ``regularity structures'' and rough-path
analysis \cite{rough_path0,rough_path}.

For SPDE and in stochastic field theory, the
use of Poisson-Kac and GPK processes provides an
interesting alternative approach in order to study these
models using a.e. differentiable stochastic perturbations (which
are definitely simpler to handle both numerically and theoretically), and
considering their Kac limit for approaching the nowhere-differentiable
case.

Let us clarify this approach with a very simple example, leaving
the analysis of physically interesting SPDE to future works.
Let  $\Omega$ be a bounded domain
of ${\mathbb R}^n$ and ${\mathcal L} : {\mathcal D}(\Omega)
\rightarrow L^2(\Omega)$ a differentiable operator, mapping
a subset ${\mathcal D}(\Omega) \subset L^2(\Omega)$ into
$L^2(\Omega)$, equipped
with suitable boundary conditions at $\partial \Omega$.
Assume that ${\mathcal L}$, equipped with the given boundary conditions,
admit a complete eigenbasis $\{ \psi_k({\bf x}) \}_{k=1}^\infty$
\begin{equation}
{\mathcal L}[\psi_k({\bf x})] = \mu_k \, \psi_k({\bf x})
\label{eq9_1_1}
\end{equation}
normalized to unit $L^2$-norm,
spanning $L^2(\Omega)$.
The simplest case is ${\mathcal L}=\nabla^2$
equipped at the boundary of the domain,  say with 
 homogeneous Dirichlet conditions.  

Let us  consider a linear SPDE, given by
\begin{equation}
\partial_t c({\bf x},t) = {\mathcal L}[c({\bf x},t)]  + b({\bf x})
\, (-1)^{\chi(t)}
\label{eq9_1_2}
\end{equation}
where $\chi(t)$ is a simple Poisson process characterized
by the transition rate $\lambda$. If ${\mathcal L}=\nabla^2$, eq. 
(\ref{eq9_1_2}) is a modified form of the Edwards-Wilkinson model
\cite{edwards,racz} of interface dynamics.
Setting $c(x,t) = \sum_{k=1}^\infty c_k(t) \, \psi_k({\bf x})$,
eq. (\ref{eq9_1_2}) reduces to the system
of stochastic differential equations for the
Fourier coefficients
\begin{equation}
d c_k(t) = \mu_k \, c_k(t) \, dt + b_k \, (-1)^{\chi(t)} \, dt
\label{eq9_1_3}
\end{equation}
where $b_k = \int_{\Omega} b({\bf x}) \, \psi_k({\bf x}) \, d {\bf x}$.
The evolution equations for the associated partial waves
$p^{\pm}(\{c_k \}_{k=1}^\infty,t)$ thus become
\begin{equation}
\partial_t p^{\pm} = -\sum_{k=1}^\infty \partial_k \left [ (\mu_k \, c_k \pm b_k
) \, p^\pm\right ] \mp \lambda \, ( p^+ -p^- )
\label{eq9_1_4}
\end{equation}
that can be solved truncating the summation up to a given integer $N$.
From eq. (\ref{eq9_1_4}) all the information on the mean field
\begin{equation}
\langle c({\bf x},t) \rangle = \sum_{k=1}^\infty \psi_k({\bf x}) \, \int
c_k \, p(\{ c_k \}_{k=1}^\infty,t) \, d {\bf c}
\label{eq9_1_4bis}
\end{equation}
 where $d {\bf c}=\prod_{k=1}^\infty d c_k$, as well as on the
correlation functions can be derived.

The problem analyzed above is fairly simple as the noise perturbation
does not depend on ${\bf x}$. It is however straightforward 
to consider space-time Poisson processes representing mollifications
of  delta-correlated stochastic perturbations both in time and in space,
which is the classical prototype of stochastic forcing 
in many problems involving SPDE.

Consider for example a one-dimensional space dimension.
Since the spatial coordinate is defined also for negative values,
the extension of  a  Poisson-Kac  process over the real
line is necessary. This can be performed, as for  the
Wiener case, by considering two independent Poisson processes
$\chi_{1^\prime}(x)$ and $\chi_{1^{\prime \prime}}(x)$,  possessing
the same transition rate $\lambda_1$, and
defined for $x \geq 0$, by introducing  the extended process $\chi_1(x)$
defined for $x \in {\mathbb R}$ as
\begin{equation}
\chi_1(x)=
\left \{
\begin{array}{lll}
\chi_{1^\prime}(-x) &  & x<0 \\
\chi_{1^{\prime \prime}}(x) & & x>0
\end{array}
\right .
\label{eq_9_add}
\end{equation}
Next
consider
process  $\chi(x,t)$, $(x,t) \in {\mathbb R} \times {\mathbb R}^+$+\{0\} defined
as 	
\begin{equation}
\chi(x,t)= \chi_1(x) + \chi_2(t)
\label{eq9_1_5}
\end{equation}
where $\chi_1(x)$ and $\chi_2(t)$ are two independent Poisson processes
characterized by transition rates $\lambda_1$ and
$\lambda_2$,  respectively, where $\chi_1(x)$ is
the extended process defined by eq. (\ref{eq_9_add}), and the SPDE
\begin{equation}
\partial_t c(x,t) = {\mathcal L}[c(x,t)] +  \alpha \, b(x) (-1)^{\chi(x,t)}
\label{eq9_1_6}
\end{equation}
$x \in {\mathbb R}$, where $\alpha>0$ is a parameter specified below.
As regards the correlation properties of the noise perturbation
one has
\begin{eqnarray}
\left \langle (-1)^{\chi(x^\prime,t^\prime)} \, (-1)^{\chi(x,t)} \right 
\rangle
& = & \left \langle (-1)^{\chi_1(x^\prime)-\chi_1(x)} \, (-1)^{\chi_2(t^\prime)
-\chi_2(t)} \right \rangle \nonumber \\
& = & e^{-2 \lambda_1 |x^\prime-x|} \, 
e^{-2 \lambda_2 |t^\prime -t|}
\label{eq9_1_7}
\end{eqnarray}
Therefore, if one sets $\alpha$ equal to
\begin{equation}
\alpha= \sqrt{ 4 \, \lambda_1 \, \lambda_2}
\label{eq9_1_7bis}
\end{equation}
the process  $\alpha \, (-1)^{\chi(x,t)}$ corresponds to a mollification
of a $\delta$-correlated process both in time and space, converging to
it in the limit $\lambda_1,,\lambda_2 \rightarrow \infty$.

The evolution equations for the Fourier coefficients of $c(x,t)$
become
\begin{equation}
d c_k(t) = \mu_k \, c_k(t) \, dt + \alpha \, b_k (-1)^{\chi_2(t)} \, dt
\label{eq9_1_8}
\end{equation}
where 
\begin{equation}
b_k = \int_{-\infty}^{\infty} (-1)^{\chi_1(x)} \, b(x) \psi_k(x) d x
\label{eq9_1_9}
\end{equation}
The expression for the random variables $b_k$
can be easily obtained by considering the dichotomous
nature of $(-1)^{\chi_1(x)}$, and the fact that the
transition instants follows an exponential distribution
defined by the transition rate $\lambda_1$.

In a similar way, nonlinear problems, as classical
stochastic fluid dynamic models (e.g. Burgers  equation),
growth models (e.g. the KPZ equation),
or the stochastic quantization of fields can be approached both numerically
and theoretically.
Once again, it is important to observe that the mollification
arising from the use of Poisson-Kac and GPK process, is not
just a mathematical artifact to regularize the structure
of a SPDE, but a way of describing physical fluctuations
possessing bounded propagation velocity, and intrinsic
 relativistic consistency.   
The extension to higher dimension is also straightforward, by
considering space-time Poisson  processes $\chi({\bf x},t)$  in ${\mathbb R}^n
\times {\mathbb R}^+\{0\}$ defined, analogously to
eq. (\ref{eq9_1_5}) as $\chi({\bf x},t)= \sum_{h=1}^n \chi_h(x_h)+
\chi_{n+1}(t)$.

\subsection{Ergodicity and $L^2$-dynamics}
\label{sec_4_2}

In this paragraph we address some issues on the 
ergodicity of Poisson-Kac and GPK processes and on some
anomalies of $L^2$-dynamics in the presence of conservative 
deterministic fields, 
considering the one-dimensional
Poisson-Kac process
\begin{equation}
d x(t) = v(x(t)) \, dt + b \, (-1)^{\chi(t)} \, dt
\label{eq9_1_10}
\end{equation}
$x \in {\mathbb R}$. This paragraph represents
a brief review  with some extensions of the 
results presented in \cite{giona_epl}.
 In one-dimensional problems, $v(x)$ can be
always regarded as a potential field deriving from
the potential $U(x)=-\int^x v(\xi) \, d \xi$.
The associated partial probability waves satisfy eqs. (A7)
of part I
where $v_\pm(x)=v(x)\pm b$. The stationary partial density
functions $p^\pm_*(x)$,  satisfy the differential equations
\begin{eqnarray}
\frac{d \left ( v_+(x) \, p^+_*(x) \right )}{ d x}
&= & -\lambda \, (p_*^+(x) - p_*^-(x))
\nonumber \\
\frac{d \left ( v_-(x) \, p^-_*(x) \right )}{ d x}& = &
 \lambda \, (p_*^+(x) - p_*^-(x))
\label{eq9_1_11}
\end{eqnarray}
from which it follows that
\begin{equation}
v_+(x) \, p_*^+(x) + v_-(x) \, p_*^-(x) = C= \mbox{constant}
\label{eq9_1_12}
\end{equation}
where the constant $C$ should be in general equal to zero because
of the regularity at infinity.
Therefore,
\begin{equation}
p_*^-(x)= - \frac{v_+(x)}{v_-(x)} \, p_*^+(x)
\label{eq9_1_13}
\end{equation}
Since by definition $v^+(x)>v_-(x)$ for $b>0$, it follows
that a stationary (positive) partial probability density may
occur solely within intervals $(a,b)$, where the conditions
\begin{equation}
v_-(x) <0 \,, \qquad v_+(x) >0 \, \qquad x \in (a,b)
\label{eq9_1_14}
\end{equation}
are satisfied. Conditions (\ref{eq9_1_14})
correspond formally to the simultaneous presence of
a forwardly propagating waves $p^+(x,t)$ and of 
a backwardly propagating wave $p^-(x,t)$.

Suppose that $v(x)$ and $b$ are such that there exists
a double sequence $x_{-,h}^*$, $x_{+,h}$, $h = -N_1,\dots,N_2$,
$N_1,N_2>0$, of abscissas
\begin{equation}
\dots < x_{+,h-1}^* < x_{-,h}^* < x_{+,h}^* < x_{-,h+1}^* < \dots
\label{eq9_1_15}
\end{equation}
such that $\{ x_{-,h}^* \}$ correspond to the nodal
points of $v_-(x)$, $v_-(x_{-,h}^*)=0$, and
 $\{ x_{+,h}^* \}$ to the nodal point of $v_+(x)$, $v_+(x_{+,h}^*)=0$.
From the above discussion, and from eq. (\ref{eq9_1_14}),
it follows that each subinterval $I_h=[x_{-,h}^*,x_{+,h}^*]$
represents an invariant interval for the partial-wave dynamics.
 If more than a single invariant interval
exists, then the stochastic dynamics (\ref{eq9_1_10}) is not ergodic,
meaning that there exists a multiplicity of stationary invariant densities,
each of which possesses compact support localized in the
invariant intervals $I_h$.

\begin{figure}[h!]
\begin{center}
{\includegraphics[height=6cm]{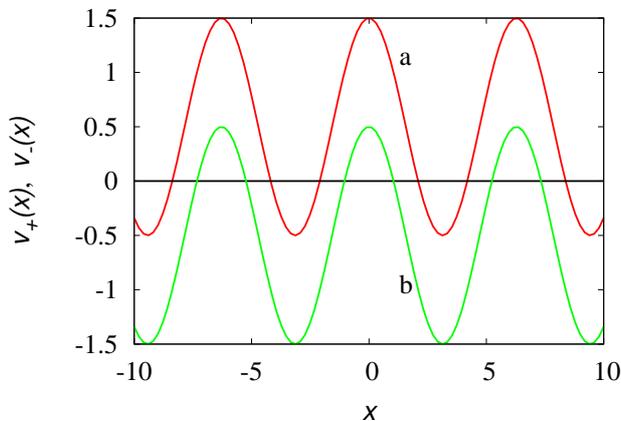}}
\end{center}
\caption{$v_+(x)$ (line a) and $v_-(x)$ (line b) for
$v(x)=\cos(x)$, $b=1/2$.}
\label{Fig22}
\end{figure}

A typical situation where invariant-density multiplicity occurs
is depicted in figure \ref{Fig22} for a sinusoidal deterministic
drift $v(x)=\cos(x)$ and $b<1$ (actually $b=1/2$). The
phenomenon of multiplicity of stationary invariant densities
disappears generically for  sufficiently 
large values of $b$ and, {\em a fortiori},  in the
Kac limit.

\begin{figure}[h!]
\begin{center}
{\includegraphics[height=10cm]{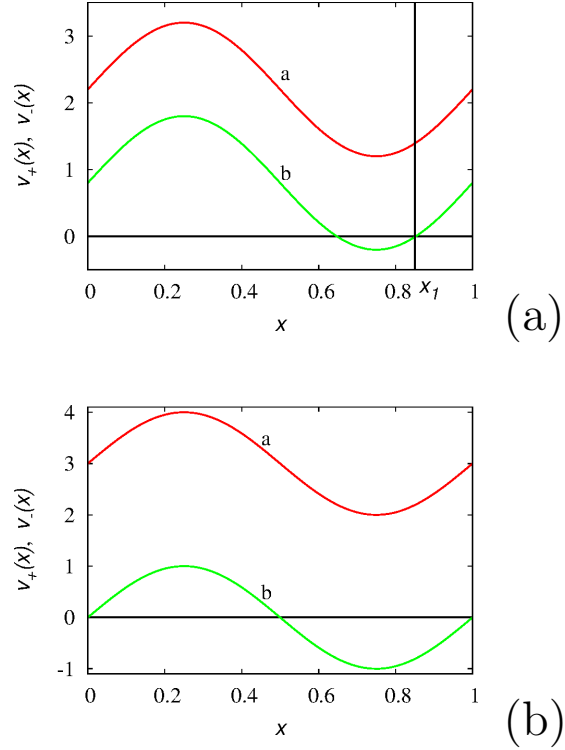}}
\end{center}
\caption{$v_+(x)$ (line a) and $v_-(x)$ (line b) for the
deterministic field $v(x)$  eq. (\ref{eq9_2_1}) at two
different values of $b$. panel (a): $b=0.7$,
panel (b); $b=3/2$.}
\label{Fig23}
\end{figure}

There is another peculiarity of Poisson-Kac and GPK processes that
should be addressed. In Section \ref{sec_2} we analyzed
the property of energy dissipation functions represented
by suitable $L^2$-norms of  the partial probability density waves in two
distinct cases: (i) in the absence of a deterministic bias,
and (ii) where ${\bf v}({\bf x})$ is solenoidal i.e., it stems
from a vector potential.
The complementary case where ${\bf v}({\bf x})$ derives from
a scalar potential, i.e., ${\bf v}({\bf x})=-\nabla \phi({\bf x})$
has not been addressed. This was not fortuitous as, in the
presence of potential velocity fields, the $L^2$-dynamics
of the partial waves may display highly anomalous and singular
properties for low values of the intensity of the stochastic
velocity $b$.
The archetype of such a singular behavior can be easily understood
by mean of the one-dimensional model (\ref{eq9_1_10}) 
defined on the unit interval $[0,1]$, and equipped with reflecting conditions
at the endpoints $x=0,1$.
As a model of the deterministic bias $v(x)$ choose, as an instance,
\begin{equation}
v(x) = \frac{3}{2} + \sin(2 \pi x)
\label{eq9_2_1}
\end{equation}
and take the stochastic velocity intensity $b$ as a parameter.
Figure \ref{Fig23} panel (a) depicts the behavior of $v_\pm(x)$
at $b=0.7$, while panel (b) refers to $b=3/2$. 
Let us analyze separately the two cases in terms of
the qualitative evolution of the partial probability waves.
With reference to the case $b=0.7$ (panel a), the interval
$[0,x^*)$, where $x^*$ is the first zero of $v_-(x)$,
$x^* \simeq 0.65$ is an escaping interval for the partial
wave dynamics: both $p^+(x,t)$ and $p^-(x,t)$ are
two progressive waves in $[0,x^*)$, so that
there exists a time instant $T^*$ such that, for $t>T^*$ and for any
initial condition, $p^\pm(x,t)=0$,  $x \in [0,x^*)$.
In the interval $[x^*,x_1]$, where $x_1$  is the second zero
of $v_-(x)$ there is the coexistence of a forwardly propagating wave
$p^+(x,t)$, and of a backwardly propagating one $p^-(x,t)$. Due to the
recombination amongst the partial waves and to the fact that the
forward $p^+$-wave propagates further towards $x>x_1$, even
this subinterval will be eventually depleted, so that,
for sufficiently long times $t$, both $p^\pm(x,t)$ 
for $x \in [x^*,x_1]$ will be arbitrarily small.
Therefore, the wave-nature of the  dynamics pushes the probability densities
 towards
the interval $(x_1,1]$. But in this region both $v_\pm(x)>0$
so that the two partial probability waves continue to
propagate forward until they reach $x=1$ where they progressively
accumulate due to the reflection conditions.

Therefore, just because of the reflecting boundary condition at $x=1$,
the unique stationary density becomes singular,
\begin{equation}
{p_*}^+(x)={p_*}^-(x)= \frac{\delta(x-1)}{2}
\label{eq9_2_2}
\end{equation}

\begin{figure}[h]
\begin{center}
{\includegraphics[height=10cm]{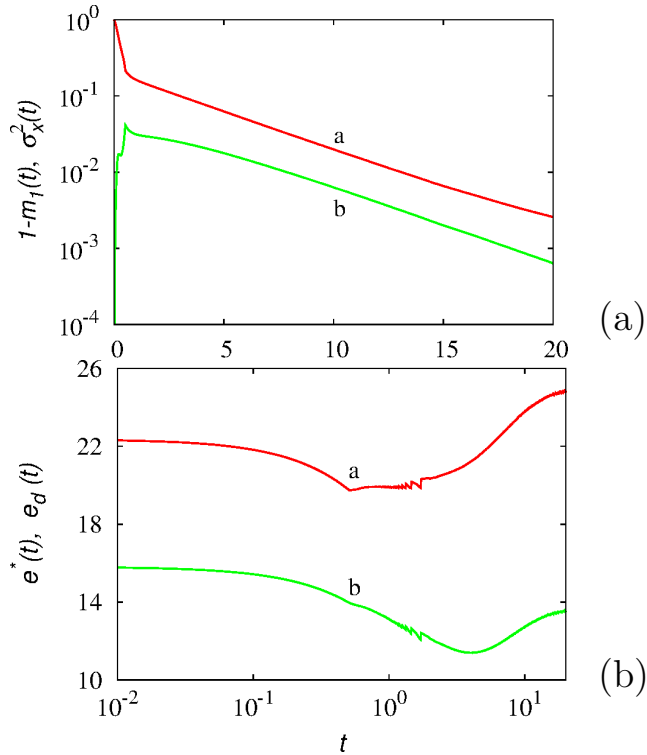}}
\end{center}
\caption{Panel (a): $1-m_1(t)$ (line a) and $\sigma_x^2(t)$
(line b) for the model problem described in the main text at $b=0.7$,
$D_{\rm eff}=1$. Panel (b): Norm dynamics for
the same problem: $e^*(t)$ (line a) and $e_d(t)$ (line b) vs $t$ (see the main
text for the definition of these quantities).}
\label{Fig24}
\end{figure}

Figure \ref{Fig24} depicts the evolution of the
moments (panel a) and of the $L^2$-norms (panel b),
 obtained from stochastic simulations
of eq. (\ref{eq9_1_10}) at $D_{\rm eff}=1$,  starting from
an initial distribution localized
at $x=0$, $p^+(x,0)=p^-(x,0)=\delta(x)/2$.
As expected from eq. (\ref{eq9_2_2}) the first-order
moments $m_1(t)$ approaches $1$ at an exponential rate
$1- m_1(t) \sim e^{-2 \lambda t}$.
The variance $\sigma_x^2(t)$ display a non-monotonic
behavior with respect to $t$, converging asymptotically to zero at the
same exponential rate.

As  regards  the $L^2$-norm depicted in panel (b),
 data have been obtained sampling a population
of $10^7$ particles using a partition of the unit interval into
$10^3$ subintervals.
In this figure $e^*(t)=||p(x,t)||_{L^2}$ and
$e_d(t)=\sqrt{{\mathcal E}_d(t)}$, where ${\mathcal E}_d(t)=(||p^+(x,t)||_{L^2}^2
+ ||p^-(x,t)||_{L^2}^2)/2$.
As expected, both these quantities admit a non-monotonic behavior and
eventually diverge for $t \rightarrow \infty$.

The occurrence of a singular impulsive invariant density
occurs for $b<b^*=3/2$ at which $v_-(1)=0$.
In this case, $b=3/2$, a unique invariant density admitting
compact non-atomic support in $[x^*,1]$, $x^*=1/2$ appears.
From  eqs. (\ref{eq9_1_11}), (\ref{eq9_1_13}), after elementary 
manipulations, the invariant density $p_*(x)$ takes the expression 
\begin{equation}
p_*(x) =\frac{A}{b^2 -v(x)} \, \exp \left [ -2 \lambda 
\int_{x^*}^x \frac{v(\xi)}{v^2(\xi) - b^2} \, d \xi \right ]
\label{eq9_2_3}
\end{equation}
where $A$ is a normalization constant.

\begin{figure}[h]
\begin{center}
{\includegraphics[height=6cm]{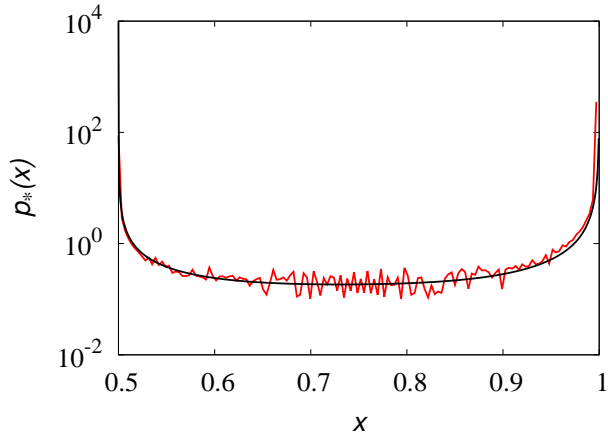}}
\end{center}
\caption{Invariant stationary probability density
 $p_*(x)$ for the  Poisson-Kac scheme
(\ref{eq9_1_10}) with $v(x)$  given by eq. (\ref{eq9_2_1})
at $D_{\rm eff}=1$, $b=3/2$. The ``noisy'' line
represents the result of stochastic simulation, the smooth
line represents eq. (\ref{eq9_2_3}).}
\label{Fig25}
\end{figure}

Figure \ref{Fig25} depicts the comparison of the
closed-form expression for the invariant  density
at $b=3/2$, $D_{\rm eff}=1$ and the results of 
stochastic simulation of eq. (\ref{eq9_1_10}).

\subsection{Ergodicity breaking in higher dimensions}
\label{sec4_3}

Ergodicity breaking occurs also in higher dimensional GPK
models in the presence of attractive and periodic potentials.
Consider a two-dimensional GPK process 
\begin{equation}
d {\bf x}(t) =  {\bf v}({\bf x}(t)) \, dt + {\bf b}_{\chi_N(t)} \, d t
\label{eqs_1}
\end{equation}
with ${\bf b}_\alpha=b \, (\cos \phi_\alpha, \sin \phi_\alpha)$,
$\phi_\alpha= 2 \pi (\alpha-1)/N$, $\alpha=1,\dots,N$,
$A_{\alpha,\beta}=1/N$, and $\lambda_\alpha=\lambda$,
$\alpha,\beta=1,\dots, N$, in the presence of
a deterministic bias ${\bf v}({\bf x})$
stemming from a potential, ${\bf v}({\bf x})= - \nabla U({\bf x})/\eta$,
which corresponds to a typical transport problem under overdamped
conditions, where $\eta$ is the friction factor.

\begin{figure}[h]
\begin{center}
{\includegraphics[height=4cm]{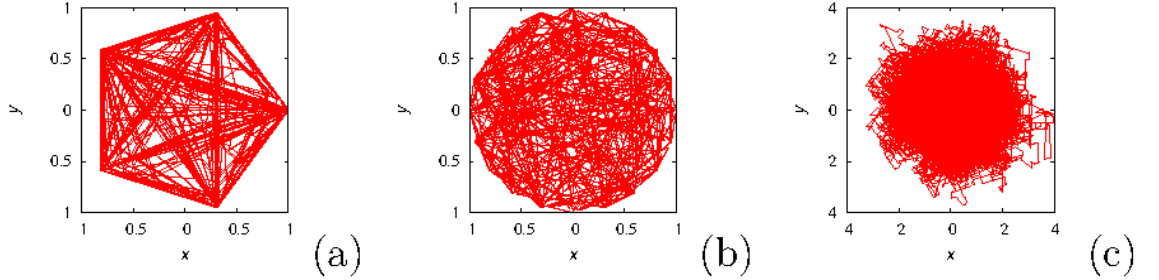}}
\end{center}
\caption{Orbits of the GPK process described in the main
text in the presence of a two-dimensional attractive harmonic potential.
Panel (a): $N=5$, $b=1$; panel (b): $N=20$, $b=1$; panel (c):
$N=5$, $b=10$.}
\label{Fig_z1}
\end{figure}

To begin with consider a harmonic, globally attractive, contribution
\begin{equation}
{\bf v}({\bf x})= - v_0 \, \left (
\begin{array}{c}
x \\
y
\end{array}
\right )
\label{eqs_2}
\end{equation}
deriving from the quadratic potential
$U({\bf x}) = U_0 (x^2+y^2)/2$, with $v_0=U_0/\eta$, and set $D_{\rm nom}=b^2/2 \lambda=1$, and $v_0=1$.
Figure \ref{Fig_z1} depicts some orbits of GPK particles
for several values of $N$ and $b$. As can be observed
particle motion is localized within an invariant region $\Omega$ of the
plane, the structure of which depends on the choice of the
stochastic velocity vectors, i.e., on $N$ and $b$.

\begin{figure}[h]
\begin{center}
{\includegraphics[height=6.0cm]{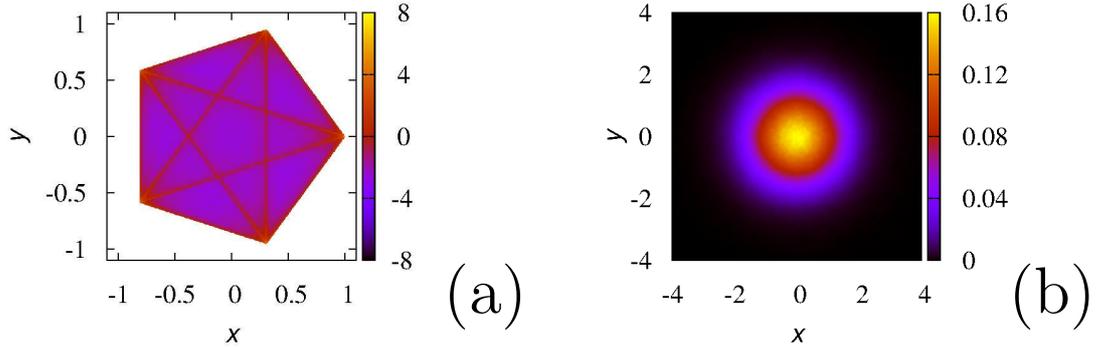}}
\end{center}
\caption{Contour plot of the compactly supported stationary invariant 
densities $p_*({\bf x})$ for
 GPK processes in the presence of a harmonic potential.
Panel (a) refers to the contour plot of $\log(p_*({\bf x}))$ at
$N=5$, $b=1$, so that $R_c=1$, panel (b) to the contour
plot of $p_*(x,y)$ at $N=5$, $b=10$, thus $R_c=10$.}
\label{Fig_z2}
\end{figure}

The structure of the invariant domain $\Omega$ can be 
derived from the condition of invariance, which dictates
\begin{equation}
\left . \left ({\bf v}({\bf x}) + {\bf b}_\alpha \right ) \cdot {\bf n}_e({\bf x})  \right |_{{\bf x} \in \partial \Omega}  \leq 0 \qquad \forall \alpha=1,\dots,N
\label{eqs_3}
\end{equation}
at the boundary $\partial \Omega$ of $\Omega$, where ${\bf n}_e({\bf x})$ is
the outwardly oriented normal unit vector at ${\bf x} \in \partial \Omega$,
and ``$\cdot$'' indicates the Euclidean scalar product.
By considering the radial symmetry of the potential, an invariant region (not
the minimal one) can be sought as a circle  of radius $R$ 
around the origin.
 Let $(r,\theta)$  be the radial coordinates. Since ${\bf v}({\bf x})=- v_0 \, r \, {\bf e}_r$, where ${\bf e}_r$ is the unit radial vector,
eq. (\ref{eqs_3}) can be expressed as
\begin{equation}
- v_0 \, R + b \, \cos(\phi_\alpha - \theta)  \leq 0
\label{eqs_4}
\end{equation}
$\alpha =1,\dots N$, $\theta \in [0,2 \pi)$, i.e.,
\begin{equation}
R  \geq \frac{b}{v_0} \, \cos(\phi_\alpha-\theta) 
\label{eqs_5}
\end{equation}
that is certainly satisfied provides that $R>R_c=b/v_0$.
The contour plots of  the stationary  invariant densities $p_*({\bf x})$
associated with two typical  GPK processes depicted in figure
\ref{Fig_z1} are shown in figure \ref{Fig_z2}.

\begin{figure}[h]
\begin{center}
{\includegraphics[height=6.cm]{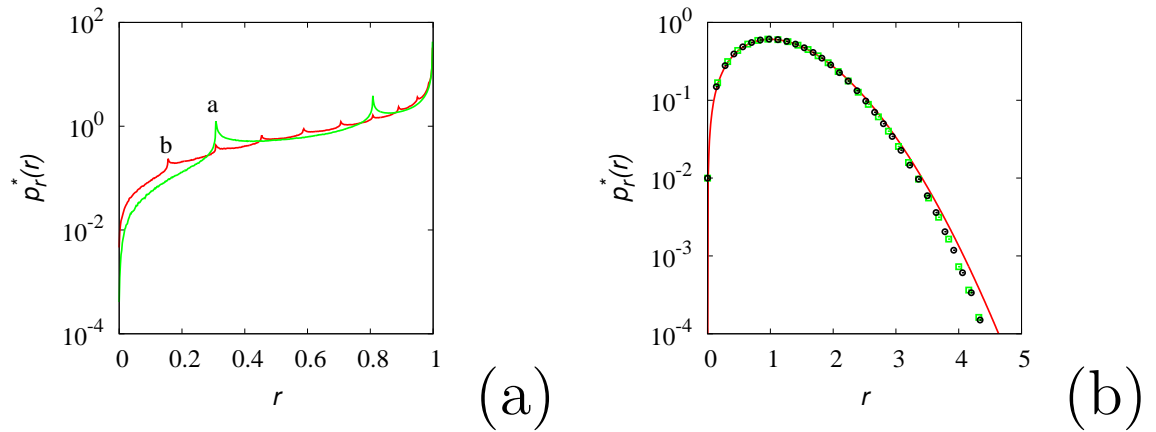}}
\end{center}
\caption{Stationary radial distribution $p^*_r(r)$ vs $r$
for GPK processes in the presence of a harmonic potential.
Panel (a) refers to $b=1$. line (a) corresponds to $N=5$,
line (b) to $N=10$. Panel (b) to $b=10$. Symbols ($\circ$)
correspond to $N=5$, ($\square$) to $N=20$. The
solid line is the Kac-limit expression for $p_r^*(r)$
eq. (\ref{eqs_7}).}
\label{Fig_z3}
\end{figure}

For low values of $b$ (panel a), the support of the invariant density
strongly depends on the geometry of the stochastic
velocity vectors (in this case, possessing a pentagonal shape, since
$N=5$). For high values of $b$, the stationary invariant density
does not depend on $\{ {\bf b}_\alpha\}_{\alpha=1}^N$, and can
be accurately approximated by the corresponding Kac-limit solution,
that in the present case provides the expression
\begin{equation}
p_*({\bf x})= A \, \exp \left [ - \frac{U({\bf x})}{ \eta \, D_0}
\right ] = A \, \exp \left [ - \frac{v_0 \, (x^2+y^2)}{2 \, D_0} \right ]
\label{eqs_6}
\end{equation}
where $A$ is a normalization constant. Figure \ref{Fig_z3}
depicts the stationary radial distribution
function $p_r*(r)$, $\int_0^\infty p_r^*(r) \, dr=1$
in the cases considered above.
For small values of $b$ (panel (a), $b=1$), $p_r^*(r)$ is 
essentially localized at the outer boundary, i.e., at $r=R_c=1$,
while for high $b's$ it practically coincides with the Kac-limit expression
\begin{equation}
p_r^*(r)= \frac{v_0}{D_0} \, r \, e^{-v_0 r^2/2 D_0}
\label{eqs_7}
\end{equation}

\begin{figure}[h]
\begin{center}
{\includegraphics[height=12cm]{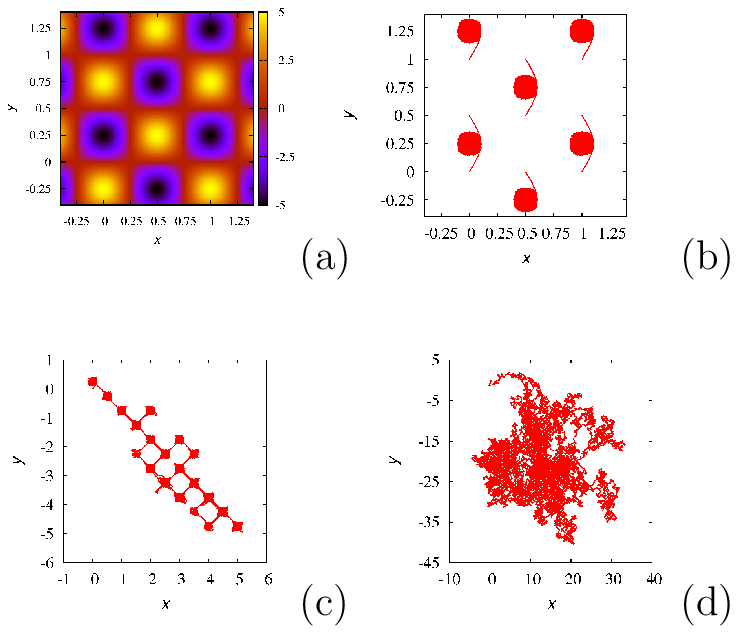}}
\end{center}
\caption{Panel (a): Contour plot of the periodic
potential (\ref{eqs_8}). Panel (b): $b=3$, orbits of the GPK
process starting from different initial positions.
Panel (c): $b=3.7$, generic orbit of the GPK process.
Panel (d): $b=10$, generic orbit of the process.
}
\label{Fig_z4}
\end{figure}

This preliminary analysis of GPK processes in radially
attractive potential is propaedeutical to the
interpretation of ergodicity-breaking phenomena
in periodic potentials.
Consider the GPK process (\ref{eqs_1}) in ${\mathbb R}^2$,
in the presence of a generic periodic potential
 ${\bf v}({\bf x})=-\nabla U({\bf x}) /\eta$, say
\begin{equation}
U({\bf x})= - \frac{U_0}{ 2 \pi} \, \cos(2 \pi x) \, \sin(2 \pi y)
\label{eqs_8}
\end{equation}
Set $\eta=1$, and $U_0=5$ for convenience, as
the analysis of ergodicity breaking is qualitatively
independent of the values attained by $\eta$ and $U_0$,
and set $N=20$, $D_{\rm nom}=1$.
Figure \ref{Fig_z4} panel (a) shows the structure of the
periodic potential (\ref{eqs_8}) considered.
Once $D_{\rm nom}$ is fixed, the only parameter of the
model is the intensity $b$ of the stochastic velocity fluctuations.
For small values of $b$, below a given threshold $b_{\rm crit}$, multiplicity
of stationary invariant measures occur,
corresponding to the presence of a countable system of
invariant regions for the GPK process located around the
local potential minima. This phenomenon is
depicted in panel (b) of figure \ref{Fig_z4} representing
some trajectories of GPK particles at $b=3$, starting from several
different initial positions that become trapped within the
invariant regions around the local potential minima. 
The critical value $b_{\rm crit}$ depends linearly
on the intensity of potential $U_0$.
For $b<b_{\rm crit}$, ergodicity breaking occurs.
Above $b_{\rm crit}$, that in the present case is approximately $b_{\rm crit} \simeq 3.59$, GPK dynamics does not display multiplicity
of localized stationary invariant measures, and ergodicity is recovered.
The qualitative behavior of the orbits above the threshold
$b_{\rm crit}$ is depicted in panels (c) and (d).
For $b \simeq b_{\rm crit}$, but above the threshold, as
in panel (c) corresponding to $b=3.7$, the orbits of GPK
particles are characterized by a ``punctured dynamics'', characterized
by longer residence times in the neighborhood of potential minima
followed by sudden jumps towards one of the nearest neighboring 
minima. Conversely, for $b \gg b_{\rm crit}$, as depicted in
panel (d) for $b=10$, GPK-particle orbits resemble   those
of a Brownian particle, and the influence of the potential
involves  the long-term dispersion properties,
the quantitative analysis of which can be recovered from the
Kac limit of the model, for sufficiently high values of $b$.

\section{Concluding remarks}
\label{sec5}

In this second part we have focused attention on the
quantitative
description of dissipation in GPK dynamics, both
in terms of energy-dissipation functions (substantially
corresponding to
$L^2$-norms of the partial probability densities) and entropies.

A correct representation of these dissipation functions should necessarily take
into account the primitive statistical formulation of the process,
based on the full system of partial probability density functions
$\{ p_\alpha({\bf x},t) \}_{\alpha=1}^N$. No consistent
energy dissipation or entropy functions can be formulated exclusively
upon the knowledge of the overall probability density function
$p({\bf x},t)$. This represents a qualitative stochastic
confirmation of the basic ansatz underlying extended thermodynamic
theories of irreversible processes.
On the other hand, the analysis of higher dimensional
GPK processes, $(n>1)$,   indicates that it is not possible to
develop a consistent thermodynamic theory of dynamic
processes possessing finite propagation velocity, by expressing
thermodynamic state variables exclusively in terms of concentrations
and their ``diffusive'' fluxes, as the whole systems of
partial probability densities (concentrations) should be
taken into account. This issue is further developed in part III.

We have also outlined another relevant application of Poisson-Kac
and GPK processes as mollifiers of space-time stochastic
perturbations in the analysis of field equations (stochastic
partial differential equations). This application
has been only sketched in Section \ref{sec_4}, and hopefully
Poisson-Kac mollification
can lead to interesting physical and mathematical results,
in the spirit of Wong-Zakai theorems and regularity-structures' theory

Moreover, we have shown that ergodicity breaking, and the
occurrence of multiple stationary invariant measures
are generic properties of GPK dynamics in higher dimensional
periodic potential, provided that the characteristic intensity $b^{(c)}$
of the stochastic velocity vectors is below a critical
threshold $b_{\rm crit}$ which depends on
the intensity and on the structure of the potential barriers.

\end{document}